\documentclass[aps,prx,reprint,onecolumn,showpacs,superscriptaddress]{revtex4-2}

\usepackage[utf8]{inputenc}
\usepackage[english]{babel}
\usepackage{float} 
\usepackage{braket} 
\usepackage{bm} 
\usepackage{xcolor} 
\usepackage{amssymb} 
\usepackage{graphicx} 
\usepackage{dcolumn} 
\usepackage{booktabs}
\usepackage{amsmath}

\begin{document}

\title{Comparative study of matrix product state/quantized tensor-train algorithms for solving time-independent partial differential equations}

\author{Paula García-Molina}
\email{paula.garcia.molina.phys@gmail.com}
\affiliation{Institute of Fundamental Physics IFF-CSIC, Calle Serrano 113b, Madrid 28006, Spain}
\author{Luca Tagliacozzo}
\affiliation{Institute of Fundamental Physics IFF-CSIC, Calle Serrano 113b, Madrid 28006, Spain}
\author{Juan José García-Ripoll}
\affiliation{Institute of Fundamental Physics IFF-CSIC, Calle Serrano 113b, Madrid 28006, Spain}

\begin{abstract}
This work presents a comparative study of new and existing optimization and diagonalization methods for solving time-independent partial differential equations (PDEs) using matrix product states (MPS) in the quantized tensor-train formalism (QTT). This study focuses on Hamiltonian equations, for which five algorithms are introduced: explicit imaginary-time evolution methods, steepest gradient descent in conventional and optimized forms, a power method, and an explicitly restarted Arnoldi method. The first five methods are engineered using a framework of limited-precision linear algebra, in which operators---i.e., the equation itself---and vectors are represented using matrix product operator (MPO) and matrix product state (MPS) formalisms, and where operator-vector multiplication and vector addition are approximated with limited resources. All methods are benchmarked using an exactly solvable PDE for a quantum harmonic oscillator in one and two dimensions over a regular grid with up to $2^{30}$ points and compared with the density matrix renormalization group (DMRG) method. Our study reveals that all MPS-based techniques exponentially outperform exact diagonalization techniques based on vectors regarding memory usage. Imaginary-time algorithms are shown to underperform any gradient descent in terms of calibration needs and costs. Finally, MPS DMRG and interpolated Arnoldi-like asymptotically outperform all other methods, including state-of-the-art vector-based exact diagonalization, with significant advantages in time and memory use.
\end{abstract}

\maketitle


\section{Introduction}

Tensor network states are a large family of quantum state representations that use moderate classical resources---time and memory---to describe complex quantum systems in scenarios of low entanglement. Tensor networks have been successfully applied in the study of quantum many-body physics\ \cite{Orus2014,  Chan2016,Bridgeman2017, Paeckel2019,ran2020}, approximating the low-energy properties of quantum Hamiltonians\ \cite{Vidal2004, Zwolak2004, Verstraete2004, White1992, White1993}, enabling the study of quantum phases of matter\ \cite{Verstraete2006, Chen2011, Schuch2011}, or the simulation of spin, bosonic and fermionic systems in multiple
dimensions\ \cite{Barthel2009, Corboz2009,tagliacozzo2009, Corboz2010, Corboz2010b, Kraus2010, Pizorn2010, Orus2014b}. The expressivity and efficiency of tensor networks have made them ideal tools to develop new \textit{quantum-inspired} algorithms---a term that describes algorithms that solve a large variety of mathematical problems with a representation that has its roots in quantum physics. Two relevant domains of application for tensor networks include optimization algorithms\ \cite{Barratt2022, Hao2022} and machine learning\ \cite{Orus2019}---e.g. unsupervised and supervised learning for the classification of images\ \cite{Stoudenmire2016, Stoudenmire2018, Liu2019, Han2018}, generative modeling\ \cite{Han2018, Cheng2019, Vieijra2022}, reinforcement learning\ \cite{Wauthier2022, Metz2022}, or the use of quantum circuits based on tensor networks to machine learning tasks\ \cite{Huggins2019}.

The first application of tensor networks involved finding the lowest energy eigenstates of some quantum-many body operator $H$. The variational algorithm for this task is known as the \textit{density matrix renormalization group} (DMRG)\ \cite{White1992, White1993}. It can be interpreted as a local optimization of the tensors in an MPS, sweeping various times over the whole system, until the expected energy converges to the lowest value\ \cite{Verstraete2004b, Schollwoeck2005, Schollwoeck2011}. A popular alternative to DMRG is solving the \textit{imaginary-time evolution} problem $\partial_t \ket{\psi}=-H\ket{\psi}$, a task which is facilitated when the Hamiltonian is local and one may apply \textit{time-evolving block decimation} (TEBD) algorithm\ \cite{Vidal2004, Zwolak2004, Verstraete2004} for a repeatedly local update of the matrix product state. Imaginary-time evolution can be upgraded through the \textit{time-dependent variational principle} (TDVP) \cite{Haegeman2011, Haegeman2016, Vanderstraeten2019, Paeckel2019}, or by combining DMRG-like methods with Taylor, Padé, and Arnoldi approximations of the evolution operator\ \cite{Manmana2005, GarciaRipoll2006}.

More recently, tensor networks have found relevant applications and demonstrated computational advantages in solving large-scale mathematical analysis problems. Of particular interest for this work is the so called quantized tensor train (QTT) format~\cite{KhoromskijOseledets2010, khoromskij2011quantics}. This is a formalism for representing discretized functions that relies on a binary encoding of the grid points, identical to the encoding of functions in quantum computers proposed by Zalka~\cite{Zalka1998} and later by Grover and Rudolph~\cite{GroverRudolph2002}. By combining this encoding with the so-called tensor train decomposition (TTs)~\cite{Oseledets2011}, the QTTs provide a highly compressed representation of functions that is amenable, among other applications, to solving partial differential equations (PDEs)~\cite{KhoromskijOseledets2010, KhoromskijOseledets2011,Dolgov2012, Khoromskij2015}. In this context, Arnoldi-like algorithms have also been employed for solving linear systems of equations, particularly using the generalized minimal residual method (GMRES)~\cite{Corona2017, Coulaud2022}. Other techniques evolve the solution on the tangent plane of the MPS/QTT manifold before projection~\cite{Uschmajew2020}, similarly to the TDVP algorithm.

The QTT formalism has been independently rediscovered in the physics world, starting from the quantum computing encoding~\cite{Zalka1998, GroverRudolph2002} and applying matrix product states (MPS)---a tensor network structure that is formally identical to tensor trains---to efficiently compress the state of the quantum register. The resulting algorithms, which are often referred to as \textit{quantum-inspired}, also include methods to solve PDEs, such as the nonlinear Schrödinger equation~\cite{Lubasch2018}, the Vlasov-Poisson~\cite{Ye2022}, the Fokker-Planck~\cite{GarciaRipoll2021}, and the Navier-Stokes equations~\cite{Gourianov2022, Kiffner2023, Holscher2024, Peddinti2024}. 

This work addresses the problem of extremal eigenvalue approximation in the MPS/QTT context for solving PDEs~\cite{GarciaRipoll2021}, where non-local Matrix Product Operators (MPOs) represent the PDEs, while MPS/QTT approximate the solutions. In this context, our work benchmarks five tensor network techniques for solving time-independent PDEs: (i) imaginary-time evolution, (ii) gradient descent methods, (iii) inverse power methods, (iv) explicitly restarted Arnoldi diagonalization, and (v) a DMRG solver. While some of these techniques have been explored in the past with QTT and TDVP projection techniques, a key contribution in this work is treat all these methods on equal footing, with a common formalism. More precisely, following Oseledets' ideas~\cite{Oseledets2011}, we start from the well-known linear algebra numerical methods for vectors, replacing in  (i)-(iv) matrix-vector multiplication and vector addition with a common finite-precision linear algebra framework based on MPOs and MPS~\cite{seemps2}. However, we improve over the original methods by introducing a state-of-the-art MPS/QTT iterative two-site update technique that optimizes the tensor decomposition~\cite{GarciaRipoll2006} to achieve the smallest tensor sizes within a given tolerance. Sweeping over pairs of sites without prescribed ranks improves precision and can solve convergence problems experienced, for instance, by single-site TDVP strategies~\cite{Yang2020}. Furthermore, simple optimizations allow us to speed up the algorithm by avoiding the construction of intermediate larger tensors, as described in Sect.~\ref{subsec:finite-precision-algebra}. In the specific case of gradient descent and Krylov methods, we improve over previous GMRES works~\cite{Corona2017, Coulaud2022}, accounting for the finite precision and loss of orthogonality,  and implementing explicit adaptive restart and extrapolation techniques that improve convergence close to the solution. Finally, this study also explores the efficient use of interpolation in MPS/QTT as a tool to accelerate convergence in the PDE solution.

The findings in this work confirm the interest of MPS tensor decompositions to treat large-scale PDEs. All tensor-network solvers lead to exponential savings in memory, enabling a simple workstation to reach problem sizes that are competitive with parallelized methods in larger supercomputers. When applicable, DMRG optimization offers the best performance, achieving exponential savings in time and converging to the lowest eigenstate in a few iterations. This is interesting, because DMRG is a method that does not respect the structure of matrix-vector algebra and rather performs an analysis of the solution at different length scales. The next best method is the Krylov-based Arnoldi algorithm, which performs best and approaches DMRG asymptotically in problems that are amenable to interpolation---e.g, when solving PDEs in grids with increasing resolution---, a situation in which the Arnoldi method benefits from a similar philosophy as DMRG. The remaining methods in order of performance are gradient descent---which can be interpreted as an Arnoldi iteration with two vectors---, the power method and, worst of all, imaginary-time evolution techniques. Finally, the optimization techniques developed in this work will find application in other realms of computational science, such as studying strongly correlated systems with MPS wavefunctions.

The structure of the manuscript is as follows. Section\ \ref{sec:Quantum-numerical-analysis} introduces the field of quantum numerical analysis, motivating the efficient encoding of partial differential equations and functions as MPO and MPS, connecting with the field of tensor trains and quantized tensor trains, and introducing the basic tools for approximate MPS-based linear algebra. Section\ \ref{sec:PDEsolvers} presents the methods to solve a time-independent multidimensional Hamiltonian PDE. These include four implementations of imaginary-time evolution (Section\ \ref{sec:ImaginaryTimeEvolution}), methods for approximate diagonalization based on gradient descent (Section\ \ref{sec:gradient-descent}) and improved gradient descent (Section\ \ref{sec:Arnoldi}), methods based on the use of the Krylov space, an Arnoldi variant with explicit restart (Section\ \ref{sec:IR-Arnoldi}) and the power iteration (Section\ \ref{sec:Power}), and finally the DMRG algorithm (Section\ \ref{sec:Power}). Section\ \ref{sec:MethodStudy}  analyzes the performance of these methods, benchmarking them against the one-dimensional quantum harmonic oscillator PDE. This study reveals the superiority of DMRG and approximate diagonalization over imaginary-time evolution. Finally, Section\ \ref{sec:Applications} compares the best MPS methods with vector-based diagonalization over a large-scale 2D problem, providing evidence of the advantages of MPS/QTT methods, as well as the best strategies to implement them. Our conclusions and outlines for future research are summarized in Section\ \ref{sec:Conclusions}.


\section{Tensor networks for mathematical analysis}
\label{sec:Quantum-numerical-analysis}

Quantum computers have been proposed as a viable platform to solve complex numerical analysis in a scalable way. For example, it has been shown how PDEs can be solved using both fault-tolerant \cite{Berry2014, Berry2017, Montanaro2016, Childs2020} and noisy-intermediate scale quantum (NISQ) algorithms\ \cite{Preskill2018, GarciaMolina2022, Lubasch2020, McArdle2019, Kyriienko2021, Knudsen2020}. These are algorithms that benefit both from an efficient compression of data into the quantum computer---the quantum register encodes highly differentiable functions with a precision that increases exponentially with the number of qubits---, and the intrinsic speed-up of quantum algorithms such as phase estimation or matrix inversion. However, despite this algorithmic progress, the fact is that contemporary quantum computers have insufficient accuracy to encode even rather simple problems\ \cite{GarciaMolina2022}. It, therefore, makes sense to look back at the progress in these quantum algorithms and understand what strategies can be reused in a classical, quantum-inspired scenario---in particular, with tensor network methods in mind.

The central idea on which the methods from this work are based is to combine an encoding of functions in quantum computers~\cite{Zalka1998, GroverRudolph2002}, with a tensor-decomposition of the resulting wavefunctions known as matrix product states\ \cite{Orus2014,Cirac2021}. As shown in Ref.~\cite{GarciaRipoll2021}, this results in an efficient and compact representation for highly differentiable multidimensional functions, which can be used to accelerate very different numerical analysis tasks---from interpolation to solving PDEs.

The tensor-network representation discussed below has been independently discovered multiple times both in the domains of quantum physics and applied mathematics, with different inspirations. As explained in Ref.~\cite{Cirac2021}, the matrix product state decomposition can be traced back to studies of strongly correlated quantum systems and quantum Markov chains~\cite{Affleck1987,Fannes1989}. MPS were later shown to be the underlying structure~\cite{Ostlund1995,Dukelsky1998} of the DMRG algorithm~\cite{White1992} for diagonalizing complex quantum many-body Hamiltonians. Finally, MPS were rediscovered under the name of tensor trains (TT) as an efficient decomposition of higher-dimensional tensors~\cite{Oseledets2011} with approximation tools similar to DMRG and MPS.

Both MPS and TT have been extended to represent functions in similar ways. First applications focused on solving the curse of dimensionality, using the factorization to correlate degrees of freedom along different spatial dimensions. This was shown to be the exact structure of certain multidimensional many-body wavefunctions~\cite{Iblisdir2007} in what is now known as the functional tensor train (FTT) expansion of a multidimensional function~\cite{Gorodetsky2019}. Such techniques have been successfully applied to solve stochastic problems~\cite{Lorenz2021, Dolgov2015, Eigel2016}, high-dimensional nonlinear PDEs~\cite{Dektor2021}, the Hamilton Jacobi Bellman equations\ \cite{Horowitz2014, Stefansson2016, Gorodetsky2018, Dolgov2019, Oster2019} and the Schrödinger equation~\cite{Hong2022}.

While the FTT expansion addresses the problem of increasing dimensions, the tensors may need to grow exponentially in size to achieve a accurate representations along each coordinate. This problem is solved by employing a tensor decomposition of the coordinates themselves~\cite{Oseledets2010}, in what is know as the quantized tensor train (QTT) algorithm~\cite{KhoromskijOseledets2010, khoromskij2011quantics}. A similar encoding has been rederived as a multidgrid expansion of functions to solve nonlinear Schrödinger equations\ \cite{Lubasch2018} and turbulence problems~\cite{Gourianov2022}, and also from a quantum-inspired point of view~\cite{GarciaRipoll2021}.

Indeed, Section~\ref{subsec:functions} shows how the quantization in the QTT representation is obtained by combining algorithms to encode functions in quantum registers with the MPS representation of quantum states. These encodings are used to represent continuous functions using qubits (Section\ \ref{subsec:functions}), and to represent potential and differential operators (Section\ \ref{subsec:operators}) as linear transformations of quantum states. Those representations are compressed using matrix product states (MPS) and matrix product operator (MPO) formalisms that are amenable to efficient manipulation in classical computers, forming the basis for a framework of MPS/QTT algorithms (Section\ \ref{sec:MethodStudy}).

\subsection{Representation of functions}
\label{subsec:functions}
The following representation of continuous functions in quantum registers is inspired by Refs.\ \cite{Zalka1998, GroverRudolph2002}. Given a $d$-dimensional function $f(\mathbf{x})=f(x_1, x_2, \dots, x_{d})$, defined over intervals $x_i \in [a_i,b_i)$ of size $L_{x_i}=|b_i-a_i|$, $i=1,\dots,d$, each dimension is discretized $x_i$ using $2^{n_i}$ points,
\begin{equation}
  \label{eq:multi-position-grid}
  x_{i,{z_i}}^{(n_i)} = a_i + z_i \Delta{x}_i^{(n_i)}.
\end{equation}
In this notation, $z_i=\in\{0,1,\ldots,2^{n_i}-1\}$ is the integer label for the grid coordinate $x_{i,{z_i}}^{(n_i)}$ of the $i$-th dimension. Each integer may be encoded in binary notation $z_i=s_{i1}s_{i2}\cdots s_{in_i}$, $s_{ir}\in\{0,1\}$, with $r=1,\dots,n_i$, using $n_i$ bits out of a quantum register. For a multidimensional function, the joined set of binary labels $\mathbf{s}=(z_{1}, z_{2}, \dots, z_{d})\in\mathbb{Z}_2^{N}$ results from grouping the states of the $N=\sum n_i$ qubits in some order. \footnote{Note that, when joining the binary labels to create a basis representative $\ket{\mathbf{s}}$, the bits of individual labels $z_i$ may be placed sequentially or they may be reordered. In the sequential order we place all bits from one coordinate together, $s_1:=s_{11},s_2:=s_{12},\ldots$. In the interleaved order, we may place the bits ordered first by significance and then by coordinate $s_1:=s_{11},\,s_2:=s_{21},\ldots$ The algorithms in this work are insensitive to this choice, but different bit orders will affect the size of the tensors involved in the representation~\cite{GarciaRipoll2021}.}Each binary number is associated to one point in the coordinate grid $\mathbf{x}_\mathbf{s}$. This enables storing the discretized function $\{f(\mathbf{x}_\mathbf{s})\}$ as the wavefunction of the quantum register, either as
\begin{eqnarray}
  \ket{f^{(N)}} = \frac{1}{\mathcal{\tilde{N}}_f^{1/2}} \sum_{\lbrace s_i\rbrace} f(\mathbf{x}_\mathbf{s})^{1/2}\ket{\mathbf{s}},
\end{eqnarray}
if $f(\mathbf{x})$ is non-negative\ \cite{Zalka1998, GroverRudolph2002} or, more generally, as
\begin{eqnarray} \label{eq:function-representation}
  \ket{f^{(N)}} = \frac{1}{\mathcal{N}_f^{1/2}} \sum_{\lbrace s_i\rbrace} f(\mathbf{x}_\mathbf{s})\ket{\mathbf{s}}.
\end{eqnarray}
Here, $\mathcal{N}_f$ is a normalization factor and Dirac's notation $\ket{\mathbf{s}}$ denotes the basis vectors in a Hilbert space $\mathbb{C}^{2^{n}}$ and their adjoints $\bra{\mathbf{s}}$.

\begin{figure}[t]
  \centering
  \includegraphics[width=1\linewidth]{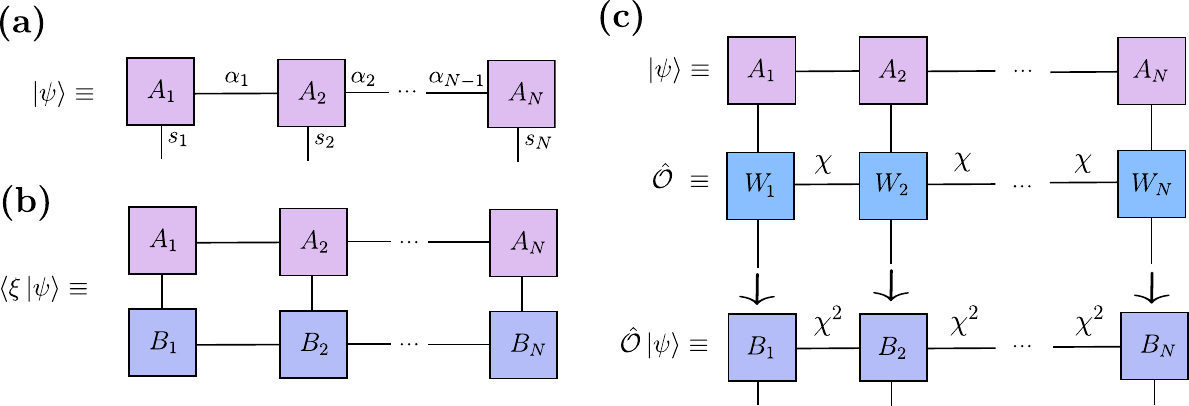}
  \vspace{-0.5cm}
  \caption{(a) Diagrammatic representation of an MPS with open boundary conditions. (b) Scalar product $\braket{\xi | \psi}$ of two states $\ket{\psi}, \ket{\xi}$ in the MPS representation. (c) Contraction of an MPS and an MPO.\label{fig:mps}}
\end{figure}

The MPS representation sketched in Figure\ \ref{fig:mps}(a) is a natural encoding for a quantum register of $N$ qubits. The exponentially many coefficients of the wavefunction are recovered by contracting a set of tensors
\begin{equation}
  \ket{f^{(N)}}=\sum_{\lbrace s_k \rbrace}\sum_{\lbrace \alpha_k \rbrace}
 (A_1)_{\alpha_1}^{s_1}(A_2)_{\alpha_{1},\alpha_2}^{s_2}\dots (A_N)_{\alpha_{N-1}}^{s_{N}}
  \times\ket{s_1}\otimes \ket{s_2} \otimes ... \otimes \ket{s_{N}},
\end{equation}
with bounded size $(A_k)_{\alpha_{k-1},\alpha_{k}}^{s_k}\in\mathbb{C}^{2\times \chi_{k-1}\times \chi_{k}}$, given by the \textit{bond dimension} or ranks $\lbrace\chi_{k-1},\chi_k\rbrace$ that depend on the entanglement content. MPS tensors $(A_k)_{\alpha_{k-1},\alpha_{k}}^{s_k}$ are in general rank-3 tensors, and only the boundary tensors $(A_1)_{\alpha_1}^{s_1}$ and $(A_N)_{\alpha_{N-1}}^{s_{N}}$ are rank-2 tensors. Provided the bond dimensions $\chi_k$ are kept small, the whole representation requires only polynomial many resources $\mathcal{O}(N\times 2\times \chi^2)$, with $\chi=\max{\chi_k}$.

\subsection{Representation of operators}
\label{subsec:operators}
The encoding of functions in the quantum register induces a similar representation for operators acting on those functions. The multiplication by a function $V(\mathbf{x})$ is a diagonal operator in the grid basis
\begin{eqnarray}
  \label{eq:potential}
  V(\mathbf{x}) := \sum_{\lbrace s_i \rbrace} V(\mathbf{x}_\mathbf{s})\ket{\mathbf{s}}\!\bra{\mathbf{s}},
\end{eqnarray}
with $\ket{\mathbf{s}}\!\bra{\mathbf{s}}$ denoting the projector onto the state $\mathbf{s}$. Differential operators of any order become diagonal operators in Fourier space
\begin{eqnarray}
  D(-i\nabla) := \bm{\mathcal{F}}^{-1} \sum_{\lbrace s_i \rbrace} D(\mathbf{p}_\mathbf{s})\ket{\mathbf{s}}\!\bra{\mathbf{s}} \bm{\mathcal{F}},
\end{eqnarray}
using the quantum Fourier transform (QFT) operator $\bm{\hat{\mathcal{F}}}$. In 1D this operator reads $\bm{\hat{\mathcal{F}}}\ket{\mathbf{r}}=\frac{1}{\sqrt{2^n}}\sum_{\mathbf{s}}\exp\left(i2\pi z_\mathbf{r}z_\mathbf{s}2^{-n}\right)\ket{\mathbf{s}},$ where $z_\mathbf{s}=\sum_i s_i 2^{n-i}$ is the binary number that results from grouping all the $n$ bits in $\textbf{s}$ (and similarly for $\mathbf{r}$). This QFT can be extended to the multidimensional scenario by a tensor product construct that takes into account the order of the bits in both $\mathbf{r}$ and $\mathbf{s}$. For a fair comparison with DMRG, this work uses finite differences as an alternative to the spectral differentiation method which avoids the cost of the QFT
\begin{align}
  \frac{\partial f(\mathbf{x})}{\partial x_i} &= \frac{f(\mathbf{x} + \Delta x_i \bm{e}_i) - f(\mathbf{x} - \Delta x_i \bm{e}_i)}{2 \Delta x_i} + O(\Delta x_i^2), \\
  \frac{\partial^2 f(\mathbf{x})}{\partial x_i^2} &= \frac{f(\mathbf{x} + \Delta x_i \bm{e}_i) - 2 f(\mathbf{x}) + f(\mathbf{x} - \Delta x_i \bm{e}_i)}{ \Delta x_i^2} + O(\Delta x_i^2),\label{eq:fdiff2}
\end{align}
one creates first and second-order derivate operators as linear combinations of displacements on the quantum register $\hat{\Sigma}^\pm$ \ \cite{GarciaRipoll2021}
\begin{eqnarray}
  \ket{\partial_{x_i}f^{(N)}} &\simeq& \frac{1}{2\Delta{x}_i}\left(\hat{\Sigma}_i^+-\hat{\Sigma}_i^-\right)\ket{f^{(N)}}, \\
  \ket{\partial^2_{x_i}f^{(N)}} &\simeq &\frac{1}{\Delta{x}_i^2}\left(\hat{\Sigma}_i^+ + \hat{\Sigma}_i^- - 2\right)\ket{f^{(N)}}.
\end{eqnarray}
The ladder operators $\hat{\Sigma}_i^+=(\hat{\Sigma}_i^-)^\dagger$ acting on the register encoding the $i$-th dimension $\hat{\Sigma}_i^+\ket{z_i} = \ket{z_i+1\mbox{ mod }2^{n_i}}$ for periodic boundary, or zeroing the $\hat{\Sigma}_i^+\ket{2^{n_i}-1} = 0$ element for open boundary ones.

\begin{figure*}[t]
  \centering
    \includegraphics[width=1\textwidth]{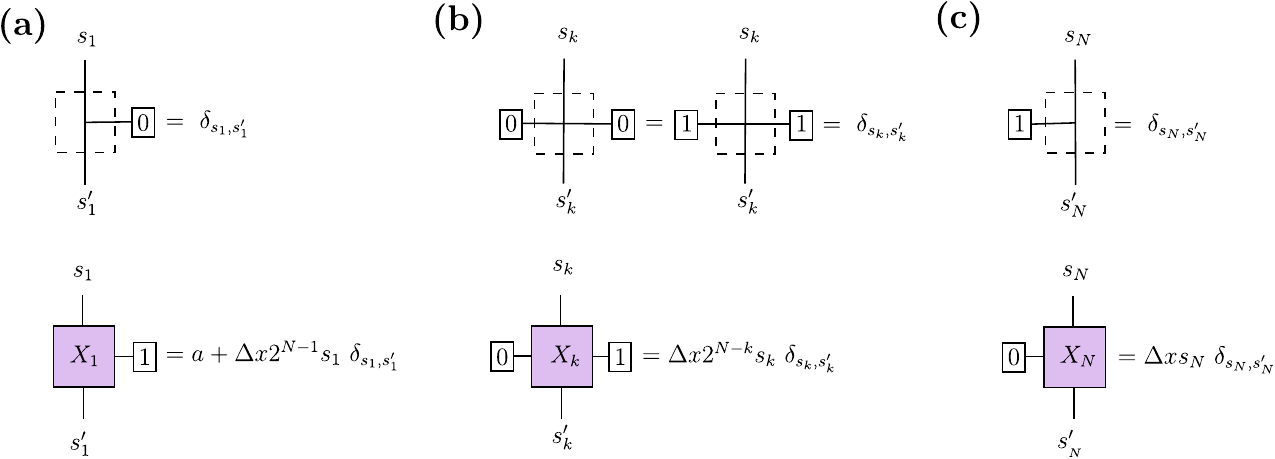}
  \vspace{-0.5cm}
  \caption{MPO elements for the one-dimensional $\hat{x}$ operator with $\chi=2$. The figure depicts the non-zero tensors' elements of the MPO. Middle tensors (b) are rank-4 tensors, while the first (a) and last (c) tensor have rank-3. If the tensor is equal to a delta function it is represented as a straight line.}
  \label{fig:x_mpo}
\end{figure*}

Just as MPS are a compressed representation for functions, MPOs are a compact representation for potentials, Fourier transforms and derivatives\ \eqref{eq:potential}-\eqref{eq:fdiff2}. Given a linear operator that has exponentially many coefficients in the qubit basis\ \cite{Chan2016}
\begin{eqnarray}\label{e4_1}
  \hat{O}=\sum_{\lbrace s_k, s_k'\rbrace}O_{s_1',...,s_{N}'}^{s_1,...,s_{N}} \ket{s_1,\dots,s_{N}}\!\bra{s_1',\dots,s_{N}'},
\end{eqnarray}
its MPO representation~\cite{Verstraete2004} is a contraction of three- and four-legged tensors,
\begin{align}\label{e4_2}
\begin{split}
O_{s_1',...,s_{N}'}^{s_1,...,s_{N}}=\sum_{\lbrace\beta_k\rbrace} & (W_1)_{\beta_1}^{s_1,s_1'}(W_2)_{\beta_1,\beta_2}^{s_2,s_2'}...(W_N)_{\beta_{N-1}}^{s_{N},s_{N}'}.
\end{split}
\end{align}
Each tensor $(W_k)_{\beta_{k-1},\beta_k}^{s_k,s_k'}$ has two physical indices, relating the qubit degrees of freedom $s_k$ and $s_k'$, and two internal indices, $\beta_{k-1}$ and $\beta_{k}$, that carry information about correlations. When these correlations are small and independent of the problem size $N$, the complete MPO requires only polynomially many components to be described $\mathcal{O}(N\times 4\times \chi^2)$, as evidenced in physical studies~\cite{Verstraete2004} and later on in the study of efficient matrix encodings~\cite{Oseledets2010}.

Computing exactly the tensors for a generic linear operator is an exponentially costly task. However, there exist simple definitions for many operators, such as the Fourier transform $\bm{\mathcal{F}}$~\cite{Chen2022,GarciaRipoll2021}, the differential operators $D(-i\nabla)$ and the $\hat{\Sigma}^\pm$ introduced above. For instance, as shown in Figure\ \ref{fig:x_mpo}, the position operator $\hat{x}$ over a given set of qubits uses tensors with bond dimension $\chi=2$. This operator can be used as a primitive to implement $V(\mathbf{x})$ or $D(-i\nabla)$. Similar representations are found for $\hat{\Sigma}^\pm$, see\ \cite{GarciaRipoll2021}. More general, a generalization of the TT-cross approximation can extrapolate some operators sampling a polynomially small set of their elements \cite{dolgov2018}.

\subsection{Finite-precision linear algebra with tensors}
\label{subsec:finite-precision-algebra}
The MPS and MPO representations are completed with four primitives to implement arbitrary linear algebra algorithms. The first operations are the scalar products $\braket{\psi|\xi}$ (Figure\ \ref{fig:mps}(b)) and $\braket{\psi|O|\xi}$. When $\psi$, $\xi$ and $O$ are expressed as tensor networks, these become contractions of quasi-2D tensor lattices. These contractions enable the computation of distances and projections and transformations of states. They are stable if the MPS and MPO are in canonical form~\cite{Cirac2021}, in which case they are only limited by the computer's finite precision.

The next two operations involve the action of an MPO onto a state $O\ket{\psi}$ (Figure\ \ref{fig:mps}(c)) and estimating a linear combination of two vectors $\alpha\ket{\psi} + \beta\ket{\xi}$. These are tasks that, when implemented naively, lead to a polynomial increase in the tensor size. A canonical strategy is to define both tasks as optimization problems, searching for the MPS that best approximates either computation
\begin{align}\label{eq:truncation1}
   \ket{\theta} &= \mathrm{argmin}_{|\theta\rangle \in \mathcal{M}_\chi} \left\Vert\ket{\theta} - O\ket{\psi}\right\Vert^2,\;\mbox{or}\\ \label{eq:truncation2}
   \ket{\theta} &= \mathrm{argmin}_{|\theta\rangle \in \mathcal{M}_\chi} \left\Vert\ket{\theta} - \sum_{l=1}^L\alpha_l\ket{\psi_l}\right\Vert^2,
\end{align}
within the space $\mathcal{M}_\chi$ of MPS with bounded resources $\chi$ (see appendiz \ref{app:truncation}).

In contrast to conventional techinques that fix the bond dimension a priori~\cite{Schollwoeck2011} and tensor-train updates that sequentially modify the tensors~\cite{Oseledets2011}, our implementation~\cite{seemps2} is based on DMRG-like collective updates of the tensors in neighboring sites---what is often known as the two-site update. More precisely, we solve equations~\eqref{eq:truncation1} and~\eqref{eq:truncation2} iteratively, finding the tensors that optimize the cost function for a pair of neighboring sites, and optimally splitting those tensors using SVD decompositions. This results in a DMRG-like iteration that proceeds back and forth along the tensor train, converging to an optimal decomposition and truncation within a predefined tolerance.

In the context of strongly correlated systems~\cite{GarciaRipoll2006}, the adaptive rank strategy has been shown to adequately propagate correlations and achieve lower bond dimensions for a prescribed accuracy. This is relevant in algorithms, such as explicit methods in Sect.~\ref{sec:ImaginaryTimeEvolution} and Arnoldi/GMRES techniques in Sect.~\ref{sec:IR-Arnoldi}, that can become unstable due to the loss of precission in each simplification step. It also addresses problems in TDVP-like algorithms, that may fail to recreate the correlations in the MPS/QTT when updating site by site~\cite{Yang2020}.

The solution of these nonlinear optimization problems enables us to implement a finite-precision linear algebra framework that lays the foundation to all other algorithms: from estimating time evolution of states under continuous equations, to implementing gradient descent or approximate diagonalization of Hermitian operators. Similar strategies have been developed in the context of DMRG\ \cite{Manmana2005} and QTT~\cite{Oseledets2011}. Compared to DMRG a crucial difference is in that the optimizations described above the states under consideration do not share any tensors with the target states, $O\ket{\psi},\,\ket{\psi}$, $\ket{\xi}$ or $\ket{\psi_l}$. This means that one may, using $L$ matrix product states of bond-dimension $\chi$, implement an optimization that in DMRG would require environments of size $L\times \chi$ and operations that scale as $L^2$ times worse, in general\ \cite{GarciaRipoll2006}. Compared to Oseledets' approach~\cite{Oseledets2011} and TDVP algorithms, another improvement is the fact that we do not need to construct the enlarged version of the linear combination QTT state. Instead, one can realize that the cost function~\ref{eq:truncation2} is a sum of scalar products $\braket{\theta|\psi_l}$ from which the optimal tensor for any set of neighboring sites may directly inferred. This will become of great importance for the practical implementation of the Hamiltonian diagonalization algorithms (Sec.\ \ref{sec:PDEsolvers}), in which linear combinations of functions are one of the costlier operations.

\subsection{Benchmark problem} \label{sec:benchmark problem}
This work will compare various algorithms that solve time-independent PDEs of Hamiltonian type, searching for the function $f(\mathbf{x})$ that satisfies
\begin{eqnarray}\label{eq:Hamiltonian-PDE}
  [D(-i\nabla)+V(\mathbf{x})]f(\mathbf{x})=E_0 f(\mathbf{x}),
\end{eqnarray}
where $E_0$ is the lowest eigenvalue of the Hamiltonian operator $H = D(-i\nabla) + V(\mathbf{x})$. The specific benchmark problem in this study is the quantum harmonic oscillator
\begin{eqnarray}
  \label{eq:ho}
  H = -\frac{1}{2}\nabla^2 + \frac{1}{2}\mathbf{x}^\dagger A \mathbf{x},
\end{eqnarray}
for which all eigenvalues and eigenstates have analytical expressions that can be accurately compared to the numerical solutions.

This study will begin with one-dimensional problems, where $A=\omega^2\in\mathbb{R}^+$ provides the lowest eigenvalue $E_0=\frac{1}{2}\omega$. The second part of the work analyzes the performance of the best methods in solving a two-dimensional harmonic oscillator, with a rotation angle $\theta$ and \textit{squeezing} factor $\sigma_{\min}/\sigma_{\max}$
\begin{equation}
  A = O^T(\theta)
  \left(\begin{array}{cc}
      1/\sigma_{\mathrm{max}}^4 & 0 \\
      0  & 1/\sigma_{\mathrm{min}}^4
    \end{array}\right)
  O(\theta),\;\mbox{with }
  O(\theta) = \left(\begin{array}{cc} \cos(\theta)  & \sin(\theta) \\
                -\sin(\theta) & \cos(\theta)\end{array}\right).
\end{equation}
For this matrix, the ground state energy is given by
\begin{eqnarray}
  E_{0} = \frac{1}{2}\left(\frac{1}{\sigma_\mathrm{max}^2} + \frac{1}{\sigma_\mathrm{min}^2}\right).
\end{eqnarray}
As seen below, despite the apparent simplicity of the problem, which is gapped, classical methods require an effort to compute the ground state $f(\mathbf{x})$ and the eigenenergy $E_{0}$ that grows rapidly with the problem discretization size. We expect to benefit form the exponential compression in the MPS representation to improve this scaling~\cite{GarciaRipoll2021}.

\section{Hamiltonian diagonalization algorithms}
\label{sec:PDEsolvers}
This section explores the redesign, using the finite-precision linear algebra above, of optimization algorithms designed to find the lowest eigenvalue of a Hermitian operator. The first methods are based on imaginary-time evolution\ \cite{Vidal2004, Zwolak2004, Verstraete2004, Zaletel2015}, an Schrödinger-like equation that converges to the lowest eigenstate of a Hermitian operator. This technique is implemented using various MPS-based explicit solvers: Euler, Runge-Kutta, and other Taylor expansions of the evolution operator. Next come methods that directly address the energy functional, minimizing it through gradient search. By upgrading a gradient descent algorithm, these optimization techniques are connected to Krylov-based methods: the power iteration and Arnoldi diagonalizations. These methods are discussed under the light of finite-precision linear algebra, with considerations to implemented them in a numerically stable way. Finally, for completeness, the DMRG algorithm\ \cite{White1992, White1993} is reviewed as a diagonalization algorithm that directly optimizes the MPS tensors given a generic MPO structure.

\subsection{Imaginary-time evolution}
\label{sec:ImaginaryTimeEvolution}
Imaginary-time evolution refers to the evolution with a Schrödinger equation in which the time variable has been replaced by an imaginary number $t\to(-i\beta)$,  $\beta\in\mathbb{R}$
\begin{eqnarray}\label{eq:Schrödinger equation}
  \partial_\beta\ket{\psi(\beta)} = -H\ket{\psi}.
\end{eqnarray}
For a non-degenerate Hamiltonian $H$ with a gapped spectrum, the normalization of the solution to this equation converges the ground state of the Hermitian operator $H$
\begin{eqnarray}
\ket{\varphi_{0}} = \lim_{\beta\to\infty} \frac{1}{\braket{\psi|\psi}}\ket{\psi(\beta)},
\end{eqnarray}
provided the initial condition $\ket{\psi(0)}$ overlaps with the eigenstate $H\ket{\varphi_0}=E_0\ket{\varphi_0}$.

Formally, equation\ \eqref{eq:Schrödinger equation} is solved by the evolution operator $U(\beta) = e^{-\beta H}$. In practice, the computation of $\ket{\psi(\beta)}$ is implemented by repeated application of a linear operator that approximates $U(\Delta\beta)$ for brief periods of time. The search for such an approximant is very relevant when the MPO representation of $H$ is efficient but the unitary operator has an exponentially large bond dimension\ \cite{Bridgeman2017}.

Since numerical analysis involves MPOs with long-range correlations and possibly multiple layers of tensors (see, for instance, the QFT MPO in\ \cite{GarciaRipoll2021}), it is relevant to seek algorithms that can be approximated using the finite-precision linear algebra techniques described above. This work compares four such methods:\\
1. \underline{Euler method.} This is an explicit, first-order Taylor approximation of the evolution, with an error $\mathcal{O}(\Delta\beta)$ and simple update with a fixed time-step $\beta_k = k \times \Delta\beta$
\begin{align}
\begin{split}
\psi_0 &= \psi(\beta_0),  \\
\psi_{k+1} &= \psi_k - \Delta \beta H \psi_k, \quad \mbox{for } k=0,1,\dots,N-1.
\end{split}
\end{align}
2. \underline{Improved Euler or Heun method.} This is a second-order, fixed-step explicit method which uses two matrix-vector multiplications to achieve an error $\mathcal{O}(\Delta\beta^2)$
\begin{equation}
\psi_{k+1} = \psi_k - \frac{\Delta\beta}{2} \left[v_1 + H (\psi_k-\Delta\beta v_1)\right],\;
\mbox{with } v_1 = H \psi_k.
\end{equation}
3. \underline{Fourth-order Runge-Kutta method.} This algorithm achieves an error $\mathcal{O}(\Delta\beta^4)$ using four  matrix-vector multiplications and four linear combinations of vectors
\begin{align}
\psi_{k+1} &=\psi_k + \frac{\Delta\beta}{6}(v_1+2v_2+2v_3+v_4),\;\mbox{with} \\
v_1 &= - H \psi_k, \nonumber \\
v_2 &= - H\left(\psi_k+\frac{\Delta\beta}{2}v_1\right), \nonumber \\
v_3 &= - H\left(\psi_k+\frac{\Delta\beta}{2}v_2\right), \nonumber \\
v_4 &= - H\left(\psi_k+\Delta\beta v_3\right).\nonumber
\end{align}
4. \underline{Runge-Kutta-Fehlberg method.} This adaptative step method\ \cite{Fehlberg1968} combines a solver with an error per step $O(\Delta\beta^4)$ with an error estimator $O(\Delta\beta^5)$ to select a $\Delta\beta$ that keeps the error below a given tolerance---starting from a good guess of the initial step size. The theoretical cost starts with six evaluations of matrix-vector multiplication, plus repetitions of evolution steps if the step size is rejected.

Higher-order methods, while approximating the imaginary-time evolution with lower errors, also require more MPO-MPS operations. Calibrating this resource and the number of integration steps is crucial to accurately determine the performance requirements of various methods. These considerations also led us to discard implicit methods---e.g., an Euler implicit formula, $(1+\Delta H\beta/2)\psi_{k+1} = (1-\Delta H\beta/2)\psi_k$---which, even though can be implemented using conjugate-gradient solvers\ \cite{GarciaRipoll2021}, have a very large and very uncontrolled cost per step.

Another important consideration of all explicit methods is stability. These methods approximate $U(\Delta\beta)$ for a short evolution step in an intrinsically unstable way. To be more precise, the eigenvalues of the approximate transformation $W_n(\Delta\beta)\simeq U(\Delta\beta)+\mathcal{O}(\Delta\beta^n)$, $\lambda(W_n)$, deviate from the contracting limit $|\lambda(W_n)/\lambda(W_0)|\nless1, n>0$. Thus, improper calibration of $\Delta\beta$ will lead to a blow-up of the evolved state and lack of convergence (see the appendix to this work).

There are alternative imaginary-time evolution techniques, such as using the Suzuki-Trotter approximation to separately apply the potential $V(\hat{\mathbf{x}})$ and the differential $D(-i\nabla)$ terms\ \cite{Leforestier1990} operators. However, this approximation is only useful if the separate application of the operators is more efficient, which is not the case for general PDEs with non-local MPO representation. Other options are second-order differencing, Chebyshev polynomial expansion, and Lanczos propagation schemes~\cite{Leforestier1990}, and the use of real-time evolution techniques\ \cite{Zwolak2004}, but the cost of these methods is comparable to the optimization algorithms introduced in following sections. Finally, while the implementation of Runge-Kutta methods to approximate the time evolution within the DMRG algorithm\ \cite{Feiguin2005} has also been studied, the advantages of a linear-algebra approach described Section~\ref{subsec:finite-precision-algebra} also applied here.

\subsection{Approximate diagonalization methods}
\label{sec:approximate diagonalization}
Imaginary-time evolution is designed to solve the imaginary-time equation. The fact that the solution converges to the ground state of the problem is a lucky and useful accident. However, there are better optimization strategies that address the energy functional associated with the equation we want to solve\ \eqref{eq:Hamiltonian-PDE}
\begin{eqnarray}\label{eq:cost functional}
  |\varphi_0\rangle = \mathrm{argmin}_{|\psi\rangle} E[\psi] = \mathrm{argmin}_{|\psi\rangle} \left(\frac{\braket{\psi| H |\psi}}{\braket{\psi|\psi}}\right).
\end{eqnarray}
Given this formulation of the problem, many strategies can engineer a trajectory $\ket{\psi(\beta)}$ that aims at decreasing $E[\psi]$, such as the steepest gradient descent, the momentum gradient descent, and the adaptative gradient algorithm (AdaGrad)\ \cite{Duchi2011}.

\subsubsection{Gradient descent} \label{sec:gradient-descent}
This method is based on updating the estimate of the solution along the direction of fastest energy decrease
\begin{eqnarray}\label{eq:optimization step}
  \psi_{k+1} = \psi_k + \Delta\beta \frac{\delta E}{\delta \psi},\mbox{ with } \frac{\delta E}{\delta\psi} = (H - \braket{H} \mathbb{I})\psi,
\end{eqnarray}
where $\braket{H} = \braket{\psi | H |\psi}$. The step $\Delta\beta < 0$ determines how far to move along the direction of the functional gradient $\frac{\delta E}{\delta \psi}$. While this update rule has been derived in areas such as MPS-based machine learning\ \cite{Gorodetsky2018b, Wang2020, Barratt2022}, the fact is that the learning rate $\Delta\beta$ does not need to be a meta-parameter of the algorithm, with delicate tuning. Instead, there exists an optimum step that results from substituting the update rule\ \eqref{eq:optimization step} in the cost functional\ \eqref{eq:cost functional}, minimizing analytically with respect to $\Delta\beta$
\begin{equation}\label{eq:gradient step}
  \Delta\beta_{-} = \frac{\braket{H'^3} - \sqrt{\braket{H'^3}^2+4\braket{H'^2}^3}}{2\braket{H'^2}^2}, \mbox{ with } H' = H - \braket{H}\mathbb{I}.
\end{equation}
Compared to the imaginary-time evolution, the steepest descent algorithm provides an automatic calibration of the solution's update, which aims directly at the ground state. Besides, the simultaneous update of the MPS/QTT tensors enables easily extending this technique to sophisticated MPOs that would be too costly to use in DMRG-like algorithms.

\subsection{Krylov subspace iterative methods}\label{sec:Krylov}
Some of the most successful modern diagonalization methods are based on the Krylov subspace. A Krylov subspace of order $L$ generated by a linear operator $H$ with dimension $N\times N$ and a vector $\psi_k$ is the vectorial subspace $\mathcal{K}_L = \mathrm{lin}\{\ket{\psi_k}, H\ket{\psi_{k}},\ldots,H^{L-1}\ket{\psi_{k}}\}$.
These methods work in this subspace to iteratively approximate the solution to the problem. Some of them are the power and the Arnoldi iteration.

\subsubsection{Gradient descent as a Krylov method}
\label{sec:Arnoldi}
The rule\ \eqref{eq:optimization step} is a particular case of a more general update that involves moving in the plane spanned by the previous solution $\psi_k$ and the derivative $\xi=H\psi_k$, which is in fact the Krylov subspace with $L=2$, $\psi_{k+1} = v_0 \psi + v_1 \xi$. In the steepest descent $v_0=1-\Delta\beta\braket{H}$ and $v_1=\Delta\beta$, but we generalize this algorithm by searching the optimal vector $\mathbf{v}^T=(v_0,v_1)$ that minimizes the total cost function. The average energy in this two-dimensional subspace has a simple expression, given by the ratio of two quadratic forms
\begin{eqnarray}
  \label{eq:2d-cost-function}
  E[\chi] = E(\bm{v}) = \frac{\bm{v}^\dagger A \bm{v}}{\bm{v}^\dagger N \bm{v}},
\end{eqnarray}
with Hermitian matrices
\begin{align}
  A &= \left(\begin{array}{cc}
         \braket{\psi|H|\psi} & \braket{\psi|H|\xi} \\
         \braket{\xi|H|\psi}  & \braket{\xi|H|\xi}
       \end{array}\right) = \left(\begin{array}{cc}
          \braket{\psi|H|\psi}   & \braket{\psi|H^2|\psi} \\
          \braket{\psi|H^2|\psi} & \braket{\psi|H^3|\psi}
        \end{array}\right),\;\mbox{and}\\
  N &= \left(\begin{array}{cc}
         \braket{\psi|\psi} & \braket{\psi|\xi} \\
         \braket{\xi|\psi}  & \braket{\xi|\xi}
       \end{array}\right)
     = \left(\begin{array}{cc}
          \braket{\psi|\psi}   & \braket{\psi|H|\psi}   \\
          \braket{\psi|H|\psi} & \braket{\psi|H^2|\psi}
        \end{array}\right).
\end{align}
The critical points of the cost function\ \eqref{eq:2d-cost-function} satisfy
\begin{eqnarray}
  \frac{\delta E}{\delta \bm{v}^*} = \frac{1}{\bm{v}^\dagger N \bm{v}}\left(A \bm{v} - E(\bm{v}) N\bm{v}\right) = 0.
\end{eqnarray}
This is a generalized eigenvalue equation
\begin{eqnarray}
  \label{eq:2d-generalized-eigenvalues}
  A \bm{v} = \lambda N \bm{v},
\end{eqnarray}
where the minimum eigenvalue $\lambda = E(\bm{v})$ gives the optimal energy for the $k$-th step, and the associated direction $\mathbf{v}$ provides the steepest descent on the plane. The generalized eigenvalue problem\ \eqref{eq:2d-generalized-eigenvalues} can be solved analytically or numerically, providing both a new estimate of the energy and a new state $\ket{\psi_{k+1}}$. While formally this method gives identical states to the gradient descent, this is often a more stable routine, thanks to the implicit normalization. 

\subsubsection{Arnoldi iteration}\label{sec:IR-Arnoldi}
The obvious improvement to gradient descent is to enlarge the Krylov basis $\mathcal{K}_L = \mathrm{lin}\{\ket{\psi_k}, H\ket{\psi_{k}},\ldots,H^{L-1}\ket{\psi_{k}}\}$ used at each step of the iteration. The result is an Arnoldi method\ \cite{Arnoldi1951} that estimates the energy functional using two matrices, $A$ and $N$, containing the matrix elements of $H$ and the identity computed in the Krylov basis. This reformulates the problem of diagonalizing an exponentially large matrix—whose size scales with the number of qubits—into a generalized eigenvalue problem within the Krylov subspace. The dimensionality of this subspace is controlled by the number of Krylov vectors, typically just 5 to 10 in practice, yielding a substantial computational advantage over direct diagonalization. Solving equation\ \eqref{eq:2d-generalized-eigenvalues} provides the coefficients $\mathbf{v}_{k+1}\in\mathbb{C}^L$ of the best approximation $\psi_{k+1}$ to the solution, with which to restart the process.

The idea of using an extended Krylov basis is not new. For instance, GMRES methods have been developed for QTT formalisms~\cite{Corona2017, Coulaud2022} using Oseledets' single-site updates. However, unlike conventional vector-based Arnoldi or Lanczos frameworks, we have to work in a scenario of limited precision introduced by the tensor-train truncation, demanding further consideration. First, one must realize that it is not possible to construct a perfectly orthogonal basis, or orthogonalize a set of existing Krylov vectors in MPS form---any linear combination of MPS is subject to some truncation and rounding errors---. Ignoring this fact and applying directly Lanczos or GMRES recurrences can lead to numerical instabilities. The answer is to keep track of the loss of orthogonalitiy by storing all scalar products of the basis in a separate matrix $N$, as we did above. 

The second consideration is that scalar products and distances are also affected by finite precision, limiting the size of the Krylov basis that can be constructed. The solution to this problem is to implement explicit restarts, stopping the growth of the Krylov basis when the condition number of the scalar product matrix $N$ exceeds a safety limit. At this point, in which we risk adding a vector that is linearly dependent with the previous ones, we can solve the generalized eigenvalue problem\ \eqref{eq:2d-generalized-eigenvalues} to get the next best guess in a smaller basis. Another strategy that improves the eigenvalue convergence is to extrapolate the next vector based on previous guesses\ \cite{Pollock2021}, using a recipe $\ket{\xi_{k+1}}=(1-\gamma)\ket{\psi_{k+1}}+\ket{\psi_k}$, with the memory factor $\gamma=-0.75$.

Lanczos and Arnoldi methods have been previously applied to MPS states in different frameworks. These methods, in combination with DMRG, are useful to compute dynamical correlations functions\ \cite{Hallberg1995}, a technique later improved by the introduction of an adaptive Lanczos-vector method\ \cite{Dargel2011, Dargel2012}. Lanczos and Arnoldi methods have also been used within the context of the variational DMRG algorithm\ \cite{Manmana2005, GarciaRipoll2006} for the evolution of one-dimensional quantum states. All these frameworks focused principally on DMRG-like methodologies, potentially requiring MPS of a larger bond dimension than the techniques presented here. On the same spirit as in this work, one must remark a complementary technique, which is the use of Chebyshev filters expansions\ \cite{Holzner2011} as iterative schemes that enable approximate diagonalization around regions of the spectrum. Unlike most common implementations, we propose to update all tensors simultaneously at each iteration. This strategy is designed to more accurately capture the global structure of the state, thereby making the method more effective for highly entangled matrix product operators (MPOs), which typically arise in the representation of general partial differential equations (PDEs). Additionally, we incorporate a variational update of the tensor dimensions (Equation~\eqref{eq:truncation1}) based on a prescribed error threshold, ensuring the accuracy and efficiency of the approximation.

\subsubsection{Power method} \label{sec:Power}
The power method is an iterative algorithm to find the largest eigenvalue $\lambda_\text{max} = \left\Vert{H}\right\Vert$ of a linear operator $H$, together with the corresponding eigenvector $(H-\lambda_\text{max})\varphi_\text{max}=0$. This algorithm starts with a random vector $\psi_0$, which for faster convergence should have a large overlap with the desired eigenvector $\varphi_\text{max}$. It progressively approximates this eigenvector via a recurrence relation
\begin{equation}\label{eq:power}
  \psi_{k+1} = \frac{H\psi_k}{\left\Vert H\psi_k\right\Vert}
\end{equation}
that actually constructs a Krylov basis.
In the limit of large number of steps, the sequence of states approaches the desired eigenvector $\lim_{k\to\infty} \psi_k = \phi_\text{max}$, with the eigenvalue computed as $\lambda_\text{max}=\left\Vert{H\phi_\text{max}}\right\Vert$. The cost of each step in this iteration is dominated by the product of the operator $H$ on $\psi_k$ and its normalization, two operations that can be implemented with matrix product states and operators.

If, as in our study, we want to find the smallest eigenvalue $\lambda_0$ of a non-negative operator $H$, the method can be generalized to operate with the shifted and inverted operator $(H-\epsilon)^{-1}$, designed so that $\left\Vert(H-\epsilon)^{-1}\right\Vert=|\lambda_0|^{-1}$. The new recurrence
\begin{equation}\label{eq:inv-power}
  \psi_{k+1} = \frac{(H-\epsilon)^{-1}\psi_k}{\left\Vert (H-\epsilon)^{-1}\psi_k\right\Vert},
\end{equation}
now involves solving the linear equation $(H-\epsilon)\xi = \psi_k$ and normalizing $\psi_{k+1} = \xi / \left\Vert{\xi}\right\Vert.$ Since we are interested in solving this equation with MPO (or combinations thereof) of large complexity, instead of implementing a DMRG-like solver, we use a conjugate gradient method---a technique that brings the power method very close to a gradient descent algorithm itself.

\section{Density matrix renormalization group} \label{sec:DMRG}

The density matrix renormalization group (DMRG) is a numerical algorithm originally developed for the study of the low-energy physics of quantum many-body physics\ \cite{White1992, White1993}. This approach was adapted to the formalism of MPS\ \cite{Verstraete2004b, Schollwoeck2005, Schollwoeck2011}, extending it with a quantum information perspective. In this point of view, the MPS is a nonlinear variational representation of the quantum state that is iteratively optimized, to minimize the expectation value of the energy in a quantum many-body problem. This paradigm has been since extended to the computation of excited states\ \cite{Chandross1999, Baiardi2022}, dynamical systems\ \cite{Jeckelmann2002}, and the real-time evolution of quantum systems via time-dependent DMRG (tDMRG)\ \cite{Daley2004, White2004, Schollwoeck2006}.

In the tensor-based form, the DMRG algorithm constructs an energy functional for a family of states $\ket{\psi[A]}$ represented as matrix product states of a given size, where the $A$ represents all the tensors $A_1,\, A_2,\ldots, A_N$ involved $E[A] = \frac{\braket{\psi|H|\psi}}{\braket{\psi | \psi}}.$ From the point of view of the tensor of a single site, this is a quotient of quadratic forms
\begin{equation}
  E[A_i] = \frac{\sum_{\alpha_n,\alpha_{n+1},i_n} \sum_{\beta_n,\beta_{n+1},j_n} A_{\alpha_n,\alpha_{n+1}}^{i_n\,*} \bar{H}_{\alpha_n,\alpha_{n+1},\beta_{n},\beta_{n+1}}^{i_n,j_n} A_{\beta_n,\beta_{n+1}}^{j_n}}{\sum_{\alpha_n,\alpha_{n+1},i_n} \sum_{\beta_n,\beta_{n+1},j_n} A_{\alpha_n,\alpha_{n+1}}^{i_n\,*} \bar{N}_{\alpha_n,\alpha_{n+1},\beta_{n},\beta_{n+1}}^{i_n,j_n}A_{\beta_n,\beta_{n+1}}^{j_n}},
\end{equation}
with tensors $\bar{H}$ and $\bar{N}$ that can be deduced from the operator $H$ and all other tensors in the quantum state $\ket{\psi}$. The trick is to interpret the tensors $A_{\beta_n,\beta_{n+1}}^{j_n}$ as representations of a quantum state $\ket{\psi^{(n)}}=\sum_{\beta_n,\beta_{n+1},j_n} A_{\beta_n,\beta_{n+1}}^{j_n}\ket{\beta_nj_n\beta_{n+1}}$ in a Hilbert space, and the tensors $\bar{H}$ and $\bar{N}$ as Hermitian operators in that space
\begin{eqnarray}
  E[\psi^{(n)}] = \frac{\braket{\psi^{[n]}|\bar{H}_n|\psi^{[n]}}}{\braket{\psi^{[n]} |\bar{N}_n| \psi^{[n]}}}.
\end{eqnarray}
Thus, the minimum of the energy with respect to a single site is given by the solution of a small eigenvalue problem $\bar{H}_n |\psi^{[n]}\rangle = \lambda \bar{N}_n  |\psi^{[n]}\rangle$,
of which we need to find the smallest eigenvalue $\lambda$.

The actual optimization is usually implemented as an iterative process, which sequentially runs through the tensors of the MPS until achieving convergence in the estimated eigenvalues and the state itself. The algorithm can be made adaptive, that is, allow the size of the tensors to grow and shrink, by operating in pairs of sites instead of single tensors, $\ket{\psi^{(n)}} \sim A_{\beta_n,\beta_{n+1}}^{j_n} A_{\beta_{n+1},\beta_{n+2}}^{j_{n+1}}\ket{\beta_{n},j_n,j_{n+1},\beta_{n+1}}$, and using the singular value decomposition to determine the optimal size for $\beta_{n+1}$ on each site.

\section{Method calibration and comparison}
\label{sec:MethodStudy}
This section presents a comparison of all methods from Section\ \ref{sec:PDEsolvers}, characterizing their performance and practical cost when solving the one-dimensional quantum harmonic oscillator PDE\ \eqref{eq:ho} with $A=1$. In order to enable the use of DMRG, the equation is discretized using a second order finite-differences method. The domain of the equation is $[-L/2,L/2)$, $L=10$, and $\Delta x = L / 2^n$, where $n$ is the number of sites of the MPS, i.e., qubits of the corresponding quantum register. The computations present in this Section and in Section\ \ref{sec:Applications} were done using the SElf-Explaining Matrix-Product-State library\ \cite{seemps2}, an open source software that contains both a backend for MPS and MPO representations, as well as all the algorithms described in this work.

The study uses four figures of merit: (i) the difference $\varepsilon = |E_0 - E_n|$ between the exact eigenvalue $E_0$ and the estimate $E_n$ using $n$ qubits, (ii) the 1-norm distance $\left\Vert\varphi_0 - \psi_{\mathrm{n}}\right\Vert_1$ between the approximate solution $\psi_{\mathrm{n}}$ and the exact ground state discretized on the same grid $\varphi_0$, (iii) the infidelity with respect to $\varphi_0$
\begin{equation}\label{eq:infidelity}
1 -F = 1 - |\braket{\varphi_0 | \psi_\mathrm{n}}|^ 2,
\end{equation}
and (iv) the standard deviation of the energy on the final state
\begin{eqnarray}\label{eq:std}
\sigma = \sqrt{\braket{\psi_\text{n}|H^2|\psi_\text{n}} - \braket{\psi_\text{n}|H|\psi_\text{n}}^2}.
\end{eqnarray}
The study also analyzes the practical cost of the MPS and MPO-MPS operations, measured as CPU time. In the case of the Arnoldi iteration\ \ref{sec:Arnoldi}, the cost varies with the number of vectors $n_v$ of the Krylov basis. Since the choice of $n_v$ is heuristic---as it is in similar state-of-the-art diagonalizers (ARPACK, Matlab, Scipy)---the study compares results for different Krylov basis size.

\begin{figure}[t]
  \includegraphics[width=1\linewidth]{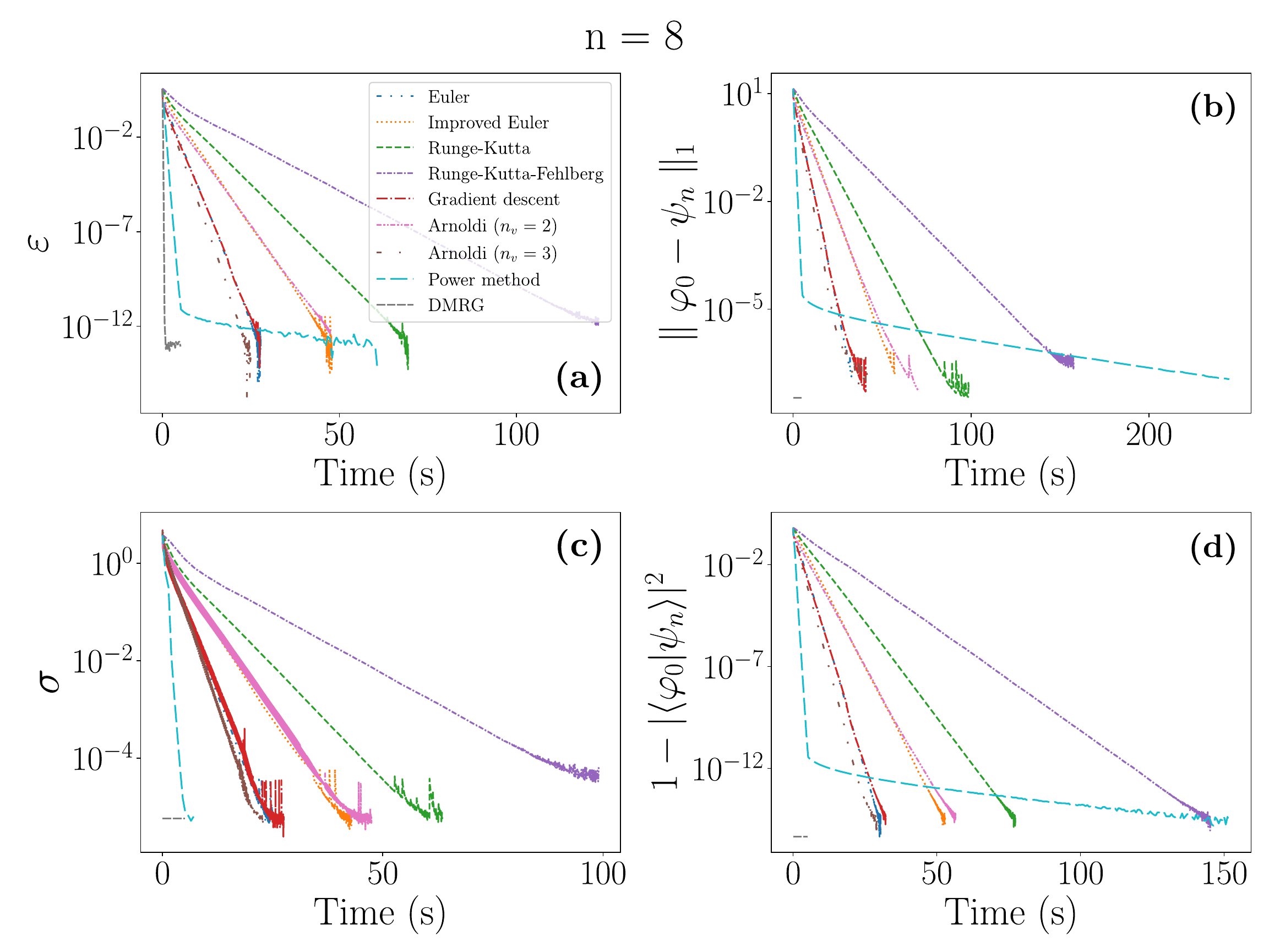}
  \vspace{-0.5cm}
  \caption{Evolution of the figures of merit versus execution time, when solving the one-dimensional quantum harmonic oscillator PDE over the interval $x\in[-L/2,L/2)$, with $L=10$, using a discretization with $n=8$ qubits and $\Delta x=L/2^n$. We plot (a) the absolute error $\varepsilon$ in the estimation of the eigenvalue, (b) the norm-1 distance, (d) the infidelity\ \eqref{eq:infidelity} with respect to the numerically exact solution, and (c) the standard deviation\ \eqref{eq:std} of the Hamiltonian over the computed eigenstate.}
  \label{fig:figures_of_merit_vs_cost}
\end{figure}

Figure\ \ref{fig:figures_of_merit_vs_cost} illustrates the four metrics of quality as a function of the execution time, when using an 8-qubit discretization and a numerically exact truncation operations of the MPS. These plots provide us the scaling of resources (time) to reach a certain error in the energy or the wavefunction. This comparison is biased towards favoring the imaginary-time evolution methods, because the time step $\Delta\beta$ is the optimal one for each method. When taking into account the time required to calibrate $\Delta\beta$ and avoid instabilities, the total CPU time increases by one order of magnitude, which renders those methods less favorable.

\begin{figure}[t]
  \centering
  \includegraphics[width=0.8\linewidth]{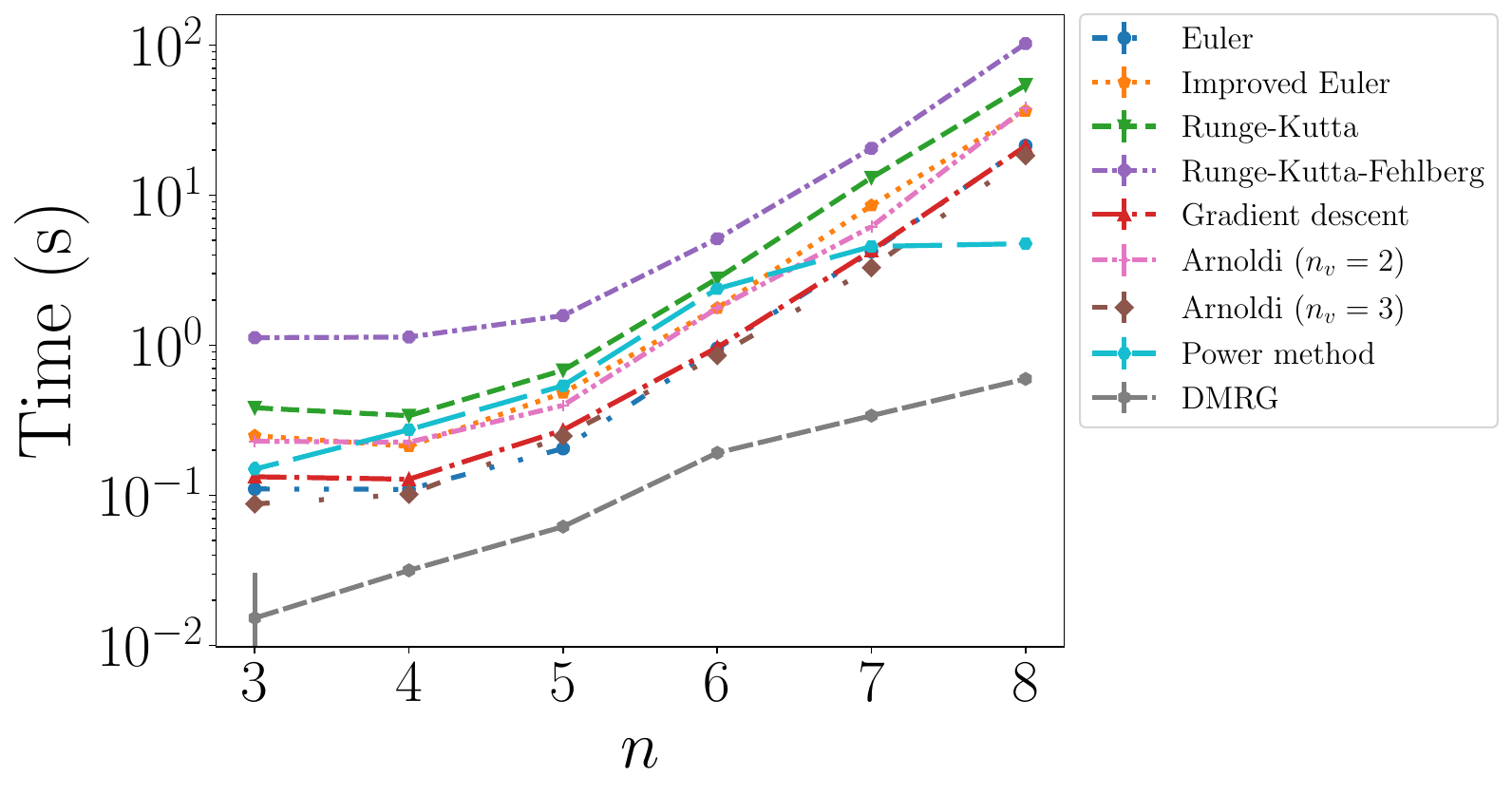}
  \vspace{-0.5cm}
  \caption{Execution time of each algorithm averaged over 10 runs, to estimate the Hamiltonian's lowest eigenvalue with an error below $10^{-10}$, when solving one-dimensional harmonic oscillator over the interval $x\in[-L/2,L/2)$, with $L=10$ and a discretization of $n$ qubits, $\Delta{x}=L/2^n$.}
  \label{fig:cost_vs_qubits}
\end{figure}

Figure\ \ref{fig:cost_vs_qubits} explores the growth in the execution time of all algorithms, for an error tolerance $\varepsilon < 10^{-10}$, as a function of the number of qubits in the discretization. All the imaginary-time evolution algorithms perform worse than the approximate diagonalization techniques, except the Euler method, which seems on par with the improved gradient descent (or Arnoldi with $n_v=2$), once more because the time does not include calibration. The conclusion of both studies is that, in general, DMRG, Arnoldi and gradient descent are the best performing methods. Furthermore, we find that the scaling of resources in DMRG has a very different power law than the alternatives (cf. Fig.~\ref{fig:cost_vs_qubits}) and, when applicable, is the best option.

\begin{figure}[t]
  \centering
  \includegraphics[width=1\linewidth]{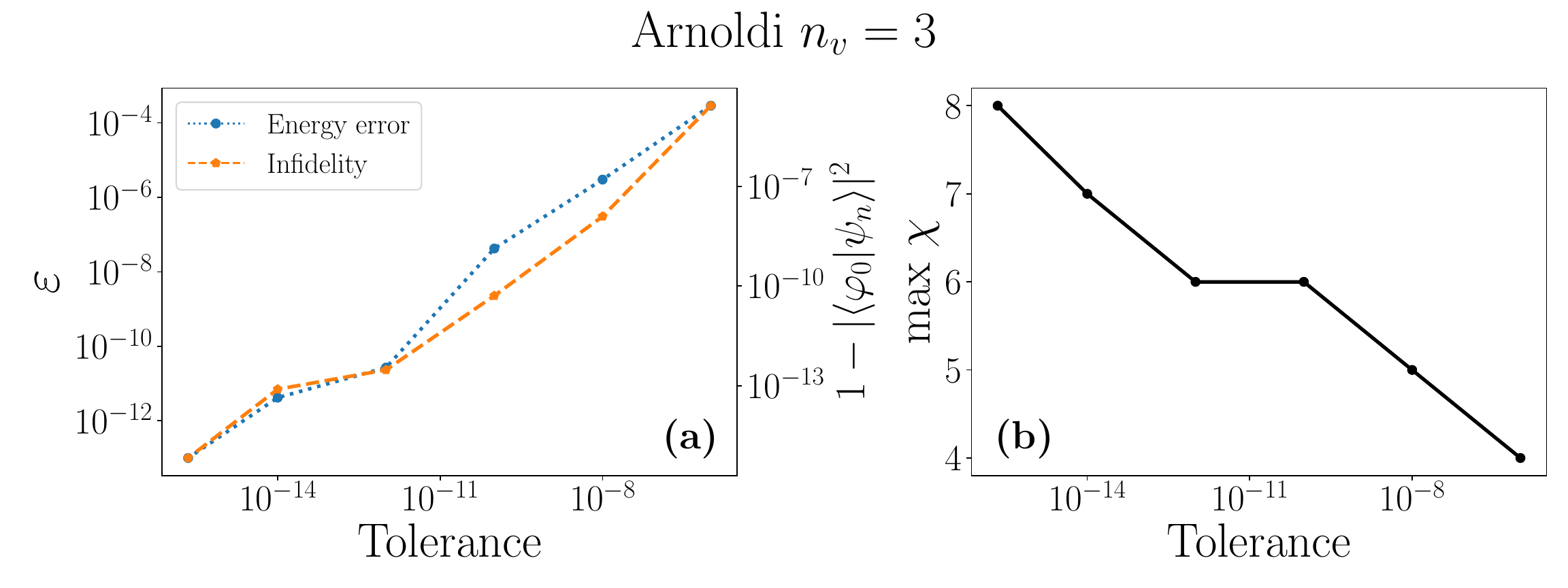}
  \vspace{-0.7cm}
  \caption{Results for different truncation tolerances in the MPS-MPS and MPO-MPS operations, for an Arnoldi diagonalization with $n_v=3$ vectors and a discretization of $n=8$ qubits. NE stands for numerically exact, indicating that the truncation tolerance is the machine precision of floating point operations. (a) Error in the estimation of the energy $\varepsilon$, (b) maximum bond dimension $\chi$ of the resulting MPS solution.}
  \label{fig:cost_tolerance}
\end{figure}

The simulations in Figures\ \ref{fig:figures_of_merit_vs_cost} and \ref{fig:cost_vs_qubits} were derived using MPS with arbitrary bond dimension, where the truncation tolerance is set to the machine's floating point precision for the squared norm of the singular values and the simplification and linear combination algorithms. Figure\ \ref{fig:cost_tolerance}(a) shows that the methods remain stable when we impose more strict truncation tolerances. A higher tolerance is accompanied by a larger error in the eigenvalue estimate (cf. Figure\ \ref{fig:cost_tolerance}(a)) and also a smaller set of tensors (cf. Figure\ \ref{fig:cost_tolerance}(b)). However, note that the tensor size $\chi$ is rather small already at the lowest tolerance, an evidence of the efficient approximation of highly differentiable functions with MPS/QTT.

\section{Benchmark: squeezed harmonic oscillator}
\label{sec:Applications}
The simulations in the previous section were intentionally small, to include all diagonalization methods, including the worst performing ones. This section lays down a larger benchmark based on the two-dimensional quantum harmonic oscillator from Section~\ref{sec:benchmark problem}, with a rotation angle $\theta=\pi/4$ and a large squeezing $\sigma_\mathrm{min} / \sigma_\mathrm{max}=0.5$, both of which increase the entanglement needs of the MPS function representation\ \cite{GarciaRipoll2021}. The comparison is now restricted to the Krylov subspace methods and DMRG, against state-of-the-art Arnoldi and Lanczos diagonalization packages based on matrix-vector multiplications---ARPACK\ \cite{ARPACK} and Primme\ \cite{PRIMME, svds_software}. For the encoding of multidimensional functions we order the qubits sequentially, first by coordinate and then by significance, following order A in Ref.\ \cite{GarciaRipoll2021}. As in Section\ \ref{sec:MethodStudy}, we set the truncation tolerance to the numerically exact machine precision according to the square norm of the singular values and the simplification and linear combination algorithms.

\begin{figure}[t]
  \centering
  \includegraphics[width=1\linewidth]{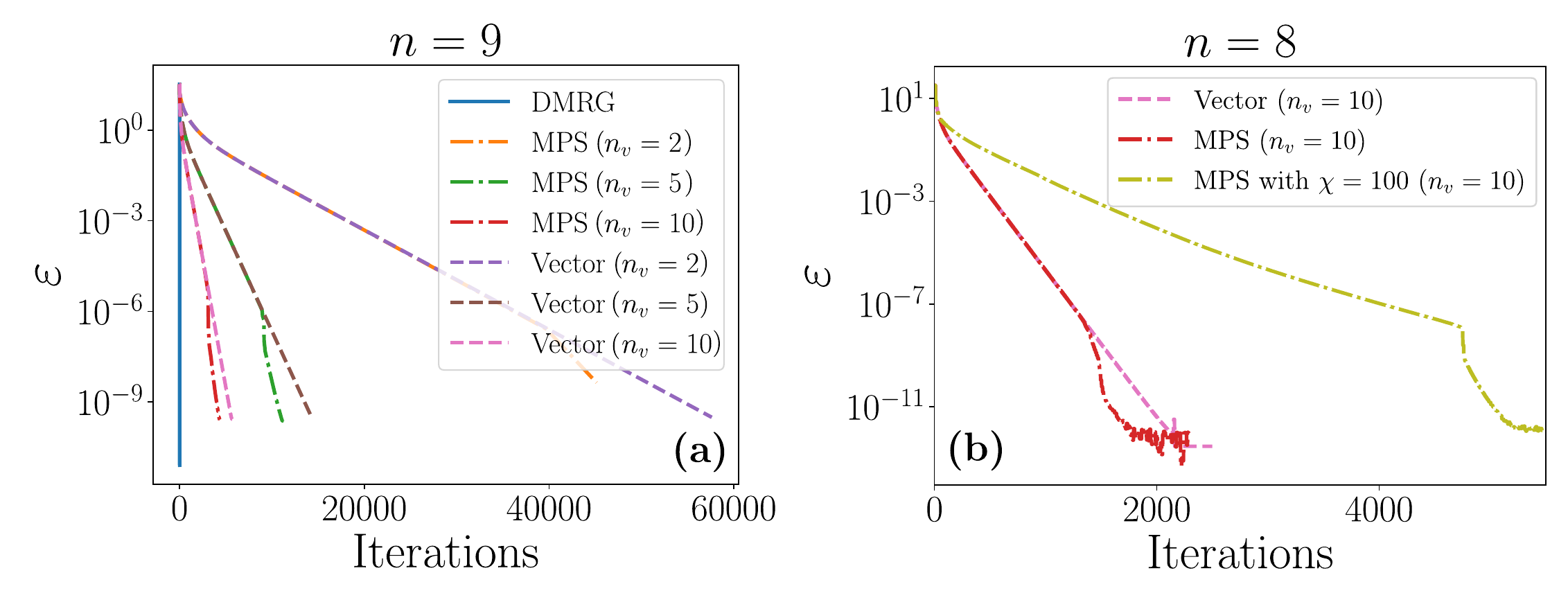}
  \caption{(a) Error in the approximation of the ground state energy $\varepsilon$ with the number of steps for the squeezed harmonic oscillator with $n=9$ qubits per dimension with arbitrarily large bond dimension for numerically exact truncation in the squared norm of the singular values and simplification and linear combination algorithms. (b) Error in the approximation of the ground state energy $\varepsilon$ with the number of steps for the squeezed harmonic oscillator with $n=8$ qubits per dimension for the Arnoldi method with $n_v=10$ for the vector and MPS implementation with arbitrarily large bond dimension and maximum bond dimension $\chi=100$.}
  \label{fig:squeezed_ho_convergence}
\end{figure}

Figure\ \ref{fig:squeezed_ho_convergence}(a) shows the decrease in the eigenvalue approximation error $\varepsilon$ as a function of the algorithmic steps, for a problem discretized with $n = 9$ qubits per dimension. All methods obtain low errors, but the Krylov-subspace methods require more steps to converge, both in the MPS and vector representations. Note how the vector and MPS representations follow more or less the same trends, except for a point at which the imprecise algebra of the MPS leads to slightly faster dissipation. Figure~\ref{fig:squeezed_ho_convergence}(b) explores the influence of setting a fixed size for the MPS tensors. Interestingly, this modification, which renders the algorithm heuristic, as it no longer follows the exact trajectory, does not affect the convergence to a low error $\varepsilon=10^{-11}$, and, while it slows it down in the number of steps required, it can speed up the practical execution of the algorithm, as the code deals with smaller tensors throughout the optimization.

\begin{figure}[t]
  \centering
  \includegraphics[width=1\linewidth]{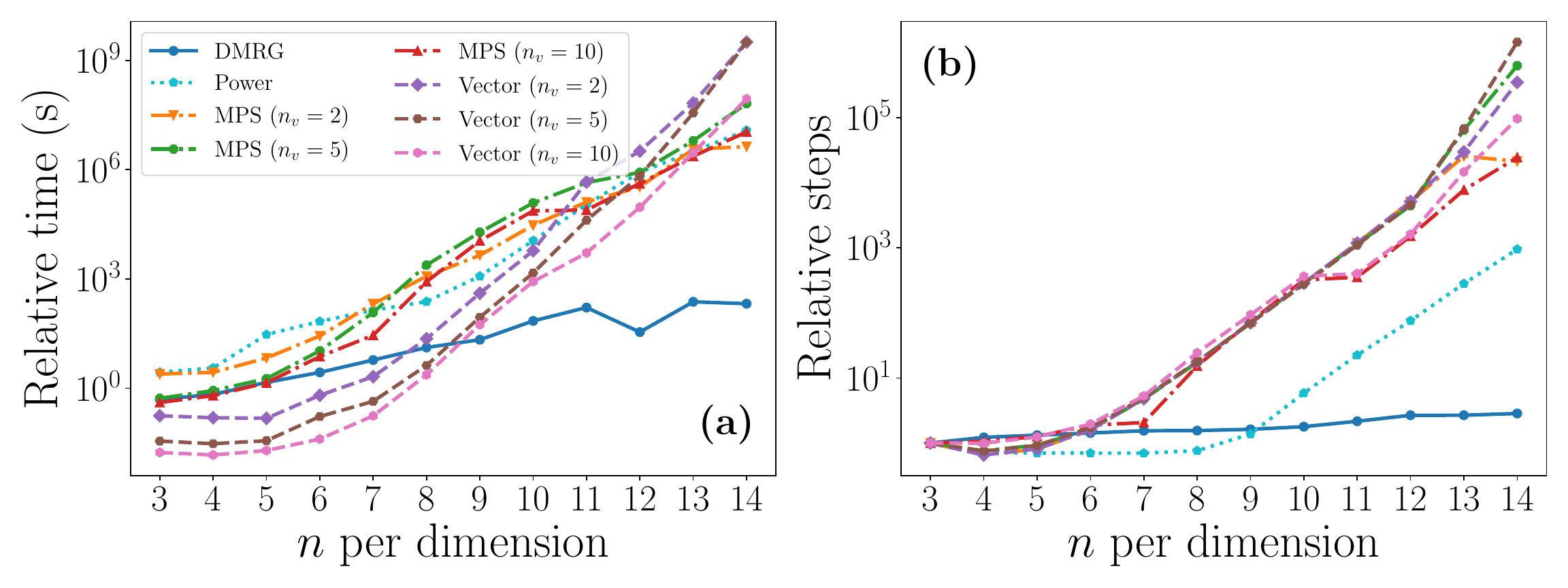}
  \vspace{-0.5cm}
  \caption{Results of the resolution of the squeezed harmonic oscillator equation\ \eqref{eq:ho}. (a) Relative time, (b) relative steps.}
  \label{fig:squeezed_ho_no_int}
\end{figure}

Figures\ \ref{fig:squeezed_ho_no_int}(a)-(b) explores the scaling of resources, by comparing them with those required in a problem with $n=3$ qubits per dimension and tensor size  $\chi=100$. Within the Krylov methods, the MPS-based algorithms exhibit a better asymptotical execution time as compared to the vector implementations. This is evidenced in the slope of the relative execution time for large problems, seen in Fig.\ \ref{fig:squeezed_ho_no_int}(a), and is fully explained by the smaller memory consumption of the MPS representation---algebraic in resources as compared to the exponential size of the vector representation.

\begin{figure}[t]
  \centering
  \includegraphics[width=1\textwidth]{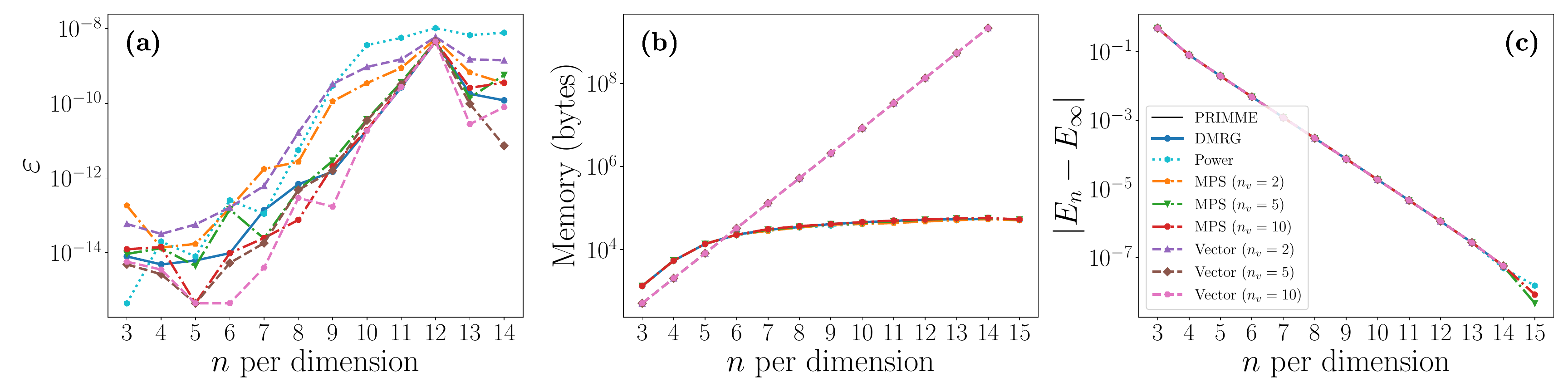}
  \vspace{-0.5cm}
  \caption{Squeezed harmonic oscillator results using finite difference interpolation \ \eqref{eq:ho}. (a) $\varepsilon$, (b) memory, and (c) theoretical error $|E_n-E_\infty|$.}
  \label{fig:squeezed_ho_int}
\end{figure}

However, even with these considerations, the DMRG algorithm still outperforms the Krylov-based methods, both in memory and in time, exhibiting an exponential advantage in the execution time. This exponential speedup can be attributed to the nonlocal approach in the DMRG optimization. Remember that DMRG solves the problem "locally" in the tensor representation, minimizing a quadratic representation of the energy functional with respect to pairs of neighboring tensors. Mathematically this means that DMRG, when sweeping from qubit 0 up to qubit $2n-1$ is solving the problem along the X and the Y directions, starting with the longest length scales first, and refining the solution within each sweep, and between consecutive sweeps.

{The understanding of DMRG's success motivated us to investigate a technique to make the Krylov-based method less prohibitively costly. This technique is based in reusing solutions, interpolating an MPS that was obtained with $n$ qubits per dimension to get a first approximation to the problem with $n+1$ qubits (see appendiz\ \ref{app:interpolation} for the interpolation algorithm). In Figure\ \ref{fig:squeezed_ho_int}(a) we observe the errors for each method. All MPS and vector methods exhibit comparable accuracies, demonstrating the capacity of MPS to obtain high precision solutions. However, now the MPS Arnoldi method becomes more competitive with DMRG in terms of execution time.}

Furthermore, this study confirms the MPS methods' better asymptotic in terms of memory resources (See Fig.\ \ref{fig:squeezed_ho_int}(b)). The vector-based algorithms exhibit an exponential growth of memory requirements and quickly exhaust our fat-node's resources, allowing us to solve problems with most $n=14$ qubits per dimension. In contrast, the MPS algorithms saturate at a modest memory size that fits in any computer, allowing us to explore sizes of up to $n=15$ qubits per dimension---i.e., a $2^{30}\simeq 10^{9}$ points grid---using around 100 Mb of practical memory (code and data) in a desktop computer.

Finally, for completeness, we show that all methods exhibit the expected precision scaling for the second-order finite-difference formulas, which scales algebraically with the discretization size $O(\Delta x^2)$ and thus exponentially with the number of qubits. In Figure\ \ref{fig:squeezed_ho_int}(c) we plot the error $|E_n-E_\infty|$ between the discrete solution with n qubits $E_n$ and the exact eigenvalue of the PDE in the continuum limit $E_\infty$. The plot includes results obtained using the Lanczos library PRIMME\ \cite{PRIMME} too. However, neither this library nor all other vector methods could provide a solution for $n=15$ qubits per dimension, which the MPS algorithms could obtain.

\section{Conclusions}
\label{sec:Conclusions}
This work has explored the solution of partial differential equations of Hamiltonian form using MPS/QTT algorithms, in which functions and operators are encoded as MPS and MPOs. The study introduced and benchmarked both new algorithms based on an efficient limited-precision linear algebra---imaginary time evolution, gradient descent, power iterations and Arnoldi techniques---with state-of-the-art DMRG. These methods have been applied to both small and large problems, exploring their asymptotic performance and practical resources.

The main conclusion is that all MPS/QTT algorithms work adequately and exhibit exponential advantages in memory over vector representations of the same problem, enabling grid sizes beyond the memory limitations of conventional techniques. Among MPS/QTT methods, imaginary-time evolution, a popular method in physics, was suprisingly less efficient than a self-calibrated gradient descent. Furthermore, Arnoldi iterations provided a significant improvement over gradient descent, in a way that was reasonably cheap and stable under the limited precision MPS algebra. Furthermore, for problems in which the PDE admits a small MPO, DMRG excels at the optimization of the state, performing exponentially better than vector-based methods also in time. When DMRG is not a suitable option, Krylov subspace methods generalize to much more complex MPOs with the use of interpolation, maintaining the exponential advantage in memory.

This work also opens many different paths for optimization and generalization. One obvious upgrade is moving from a state-of-the-art Python implementation, our SeeMPS library, to a C++ implementation. Early results show a 10x speedup that already makes the methods competitive with Arpack/PRIMME at even smaller sizes. Another obvious speedup would come from parallelizing both the MPS linear algebra operations and the construction of the Krylov basis in the Arnoldi methods. These algorithms may also be improved in terms of precision and stability, and extended to other tasks, such as diagonalization in other areas of the spectrum, computation of excited states, or the inclusion of symmetries. The precision of the MPS/QTT representation may also be enhanced, replacing finite differences with Fourier interpolation\ \cite{GarciaRipoll2021} and an exponentially more efficient encoding of derivative operators\ \cite{GarciaMolina2022}.   Finally, as described in the appendiz\ \ref{app:source-PDEs}, the techniques put forward in this work can be extended to other PDEs, such as equations with sources, via suitable reformulations that convert those equations into optimization problems.

Finally, given the good performance exhibited in this work, we believe that these diagonalization methods will become useful also in the study of many-body physics problems, joining other techniques used for long-range interactions, such as Chebyshev expansions\ \cite{Wolf2014}, the generalized TDVP algorithm\ \cite{Koffel2012, Haegeman2016}, the MPO $W^{I,II}$ method\ \cite{Zaletel2015}, or the variational uniform matrix product state (VUMPS)\ \cite{Zauner-Stauber2018} algorithm, that combines the DMRG and MPS tangent space concepts.

While MPS/QTT methods have shown promise, their effectiveness remains largely heuristic---albeit with plausible convergence and error checks. The question of why and when these algorithms will succeed, the tensor sizes for different functions and operators, the stability under tensor truncation errors, and the resilience to numerical precision are open research questions. However, recent work \cite{Jobst2024} demonstrates that functions with rapidly decaying Fourier spectra can be efficiently encoded as MPS/QTT, offering a promising direction for future research and applications.

\section*{Acknowledgments}
This work was funded ``FSE invierte en tu futuro" through an FPU Grant FPU19/03590 and by {MCIN/AEI/10.13039/ 501100011033}. JJGR and PGM acknowledge support from Proyecto Sinérgico CAM 2020 Y2020/TCS-6545 (NanoQuCoCM),  Spanish project PID2021-
127968NB-I00 funded by MCIN/AEI/ 10.13039/501100011033/FEDER, UE, and CSIC Interdisciplinary Thematic Platform (PTI) Quantum Technologies (PTI-QTEP+). JJGR acknowledges support by grant NSF PHY-1748958 to the Kavli Institute for Theoretical Physics (KITP). LT aknowledges the ``Plan Nacional Generaci\'on de Conocimiento'' PGC2018-095862-B-C22. The authors also gratefully acknowledge the Scientific computing Area (AIC), SGAI-CSIC, for their assistance while using the DRAGO Supercomputer for performing the simulations, and Centro de Supercomputación de Galicia (CESGA) for access to the supercomputer FinisTerrae.

The authors declare no conflict of interest.

\appendix
\section{MPS algebra}
\label{app:truncation}
In this work, we implement many algorithms using MPS and MPOs. We consider MPS as vectors within an algebra, constituted by the minimum set of operations to implement any algorithm based on the application of quantum operators on quantum states and the linear combination of quantum states.

The contraction of an MPO with an MPS increases the bond dimension of the resulting MPS, and for large bond dimensions, this operation becomes very costly, as the number of coefficients of the MPS increases quadratically with the bond dimension\ \cite{Orus2014}. Thus, to efficiently apply quantum operators on quantum states, truncation algorithms have been proposed to truncate the bond dimension of the tensors of the MPS, while still representing the same quantum state up to a certain error. The simplest approach is to directly truncate the Schmidt coefficients while performing the SVD\ \cite{Paeckel2019}. More stable and precise approaches are based on a variational truncation\ \cite{Paeckel2019}.

In our MPS methods we implement a two-site simplification algorithm to approximate an MPS quantum state $|\psi\rangle$ with bond dimension $\chi_\psi$ by projecting it in the subspace of MPS with bond dimension $\chi_\phi$, {$\mathrm{MPS}_{\chi_\phi}$}, such that $\chi_\phi<\chi_\psi$. The resulting MPS $|\phi\rangle\in \mathrm{MPS}_{\chi_\phi}$ is the state that minimizes the distance $d(\psi,\phi)$
\begin{align}
  \phi &= \mathrm{argmin}_{\phi \in \mathrm{MPS}_{\chi_\phi}} d(\psi,\phi),\\ \mbox{with }
  \label{eq:MPS distance}
d(\psi,\phi)&=\left\Vert\psi-\phi\right\Vert^2 =\braket{\psi|\psi}+\braket{\phi|\phi}-\braket{\psi|\phi}-\braket{\phi|\psi}.\nonumber
\end{align}
This is a bilinear function with respect to any tensor in $\ket{\phi}$, leading to a functional that can be efficiently optimized via an iterative algorithm.

\begin{figure}[t]
  \centering
  \includegraphics[width=0.75\linewidth]{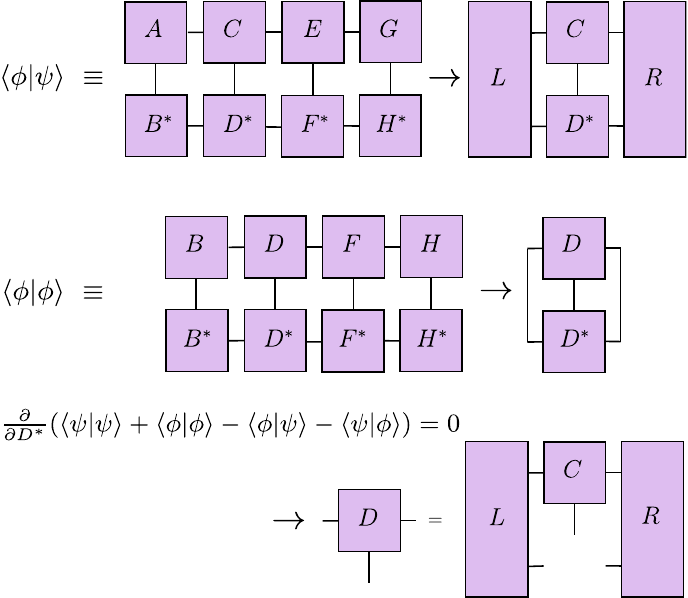}
  \caption{Diagrammatic representation of one step of the simplification algorithm for MPS.}\label{fig:mps simplification}
\end{figure}

In this algorithm, we locally optimize the MPS to minimize the distance\ \eqref{eq:MPS distance} with respect to a site $D$ using its canonical form. The minimization condition is
\begin{eqnarray}
  \frac{\partial}{\partial (D)_{\alpha\beta}^{i}}d(\psi,\phi) = \frac{\partial}{\partial (D^*)_{\alpha\beta}^{i}}d(\psi,\phi)=0.
\end{eqnarray}
Then we can compute
\begin{eqnarray}
  \frac{\partial}{\partial (D^*)_{\alpha\beta}^{i}}\langle\psi|\phi\rangle = U^{i}_{\alpha\beta}, \quad \frac{\partial}{\partial (D^*)_{\alpha\beta}^{i}}\langle\phi|\phi\rangle = (D)^{i}_{\alpha\beta},
\end{eqnarray}
where $U$ is constructed from the contraction of the left environment $L$, the right environment $R$ and $C$, and this leads to the approximation of $\phi$ for that given iteration,
\begin{eqnarray}
  (D)_{\alpha\beta}^{i} = U_{\alpha\beta}^{i}.
\end{eqnarray}
A diagrammatic representation of one step of the algorithm is depicted in Figure\ \ref{fig:mps simplification}. In practice, we use a two-site local optimization algorithm by contracting two neighboring sites, as it adapts the bond dimension at each step to achieve greater accuracy and stability. Both algorithms are implemented in the same way, by expressing the solution $(D)_{\alpha\beta}^{i} = U^i_{\alpha\beta}$ as an antilinear form that maps the local tensor $(D)^i_{\alpha\beta}$ of $|\phi\rangle$  to the scalar product of $\langle \phi|\psi\rangle$. With this algorithm, we decrease the bond dimension of the MPS while maintaining the precision up to a certain tolerance. This allows us to avoid the exponential increase in the application of quantum operators as MPO.

\begin{figure}[t]
  \centering
  \includegraphics[width=0.75\linewidth]{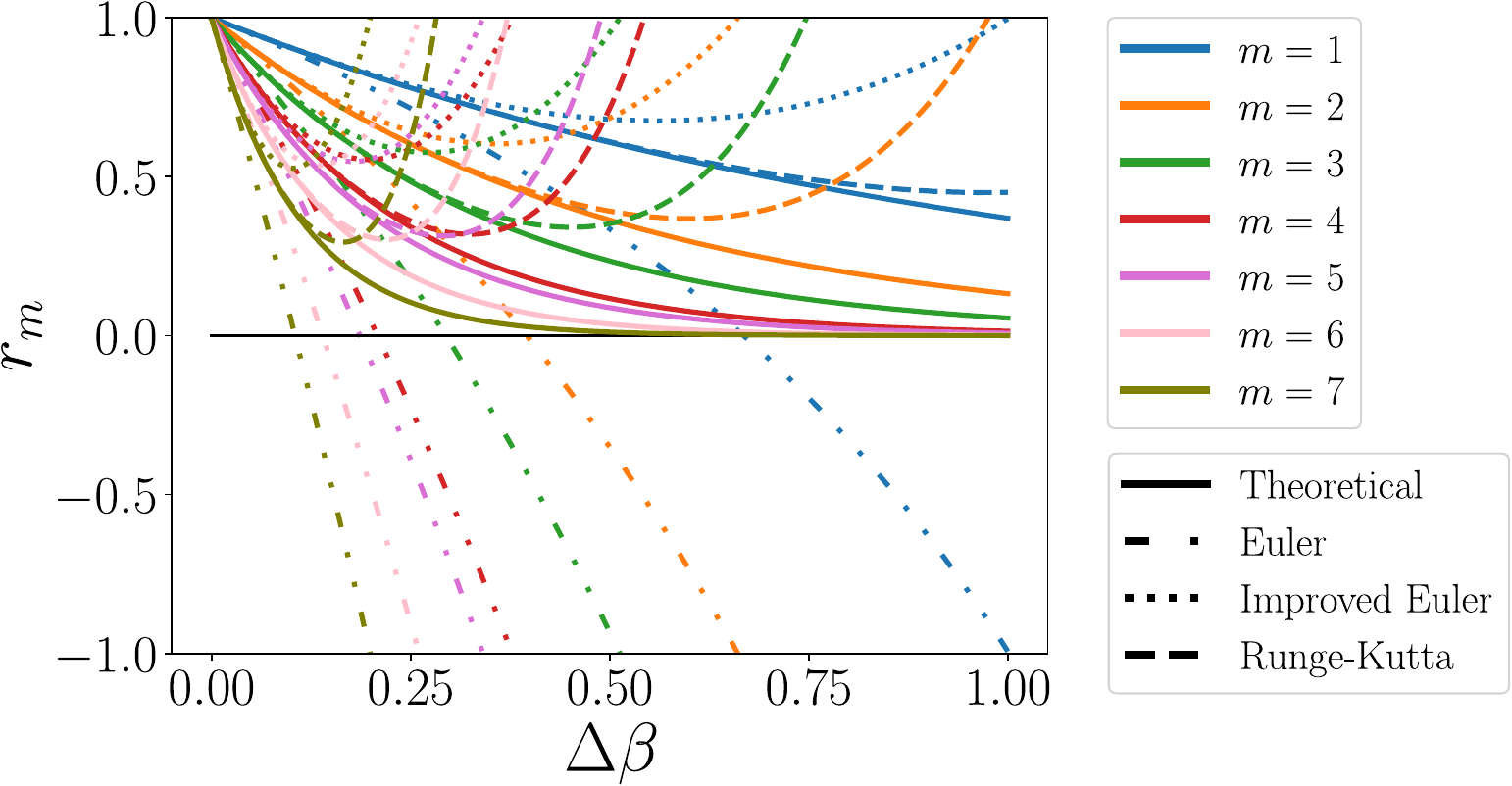}
  \caption{Contraction ratio $r_m$ of the one-dimensional quantum harmonic oscillator \eqref{eq:ho} 3-qubit discretization for $\Delta\beta \in [0,1]$. The different line styles correspond to the used methods: theoretical evolution (solid), Euler (dash-dotted), improved Euler (dotted), and Runge-Kutta (dashed).}
  \label{fig:contraction ratio}
\end{figure}

We can extend this algorithm to the approximation of a linear combination of states, i.e., to solve the problem

\begin{eqnarray}
  \mathrm{argmin}_{\phi\in\mathrm{MPS}}\left\Vert\phi- \sum_{l=1}^{L}\alpha_l\psi_l\right\Vert^2.
\end{eqnarray}
In this case we have an antilinear form for every scalar product $\langle\phi|\psi_n\rangle$, and the solution is a weighted linear combination of the solution for each state
\begin{eqnarray}
  (D)^i_{\alpha,\beta} = \sum_{n=1}^{L}\alpha_n U_{\alpha\beta}^{(n)i}.
\end{eqnarray}

\section{Time step limitation in Runge-Kutta methods} \label{app:step-size}

In the imaginary-time evolution Runge-Kutta methods in Section\ \ref{sec:ImaginaryTimeEvolution}, correctly choosing the step size $\Delta\beta$ is key to optimize the convergence to the ground state. This convergence is determined by the contraction ratio $r_m$,

\begin{eqnarray} \label{eq:contraction ratio}
  r_m = \frac{\lambda_m(\beta,E_m)}{\lambda_0(\beta,E_0)}, \quad m > 0,
\end{eqnarray}
where $\lambda_n(\Delta\beta,E_n)$ is the eigenvalue of the Runge-Kutta method for the $n$th energy level and step size $\Delta\beta$. The smaller this value, the faster the component of the corresponding $m$-th energy level goes to zero. The form of the contraction ratio depends on how the imaginary-time evolution method approximates the evolution operator, so the optimum $\Delta\beta$ varies for each numerical method. To study this, let us plot the contraction ratios of the energy levels corresponding to the quantum harmonic oscillator \eqref{eq:ho} 3-qubit discretization for a range of values of $\Delta\beta$. We represent them for $r_m \leq |1|$, as outside this interval convergence is not assured. We observe that higher-order methods approximate better the exact theoretical evolution, especially for the smallest $\Delta\beta s$, as expected due to the smaller global error associated with them. However, although the Euler method fails to reproduce the behavior of the evolution for larger values of $\Delta\beta$, it can achieve smaller contraction ratios and consequently faster convergence.

Not only the step size $\Delta\beta$, but also the initial state $\psi(\beta_0)$, plays a key role in the convergence. For $c_0 = \langle \varphi_0 | \psi(\beta_0)\rangle = 0$, the state cannot evolve to the ground state, so we need to make sure that the initial state has $c_0 \neq 0$. In addition, if the initial state has some $c_m =0, \ m > 0$, their corresponding contraction ratios will not be considered for the computation of $\Delta\beta_\mathrm{opt}$. Thus, the optimum step size will depend on the initial state.

Figure~\ref{fig:contraction ratio} shows the contraction ratios $r_m \leq |1|$ of the energy levels corresponding to the quantum harmonic oscillator~\eqref{eq:ho} 3-qubit discretization for a range of values of $\Delta\beta$. Higher-order methods provide a more accurate approximation of the exact imaginary-time evolution for small step size values, as expected. Lower-order methods, such as Euler, can achieve smaller contraction ratios and faster convergence to the ground state for larger $\Delta \beta$ than higher-order methods, which better approximate the imaginary evolution path. This indicates the possibility for other paths better suited for energy minimization, like those in Sections~\ref{sec:approximate diagonalization},~\ref{sec:Krylov} and~\ref{sec:DMRG}. 

In general, the spectrum of the problem that we are aiming to solve is unknown, so we cannot compute the contraction ratios or the $c_n$ coefficients. In these cases, we will use a minimization algorithm that finds the optimum step size according to a certain figure of merit acting as cost function, in our case the number of steps to achieve a certain error $\varepsilon$ in the approximation of the energy. This method is costly since requires as many executions as optimization steps. However, it is a better alternative than simple inspection in fewer executions of the numerical method, as optimization provides a closer approximation to the optimum step size.

\section{Interpolation}
\label{app:interpolation}

We have described a discrete approximation of continuous functions and operators. We can increase its precision using interpolation. For functions that meet the requirements for the spectral method, we can arbitrarily increase this precision---up to $O(e^{-r2^n}), $ for analytic functions, where $r$ is a problem-dependent constant\ \cite{GarciaMolina2022}---using Fourier interpolation. This technique reconstructs the original continuous, bandwidth-limited, infinitely differentiable function from the momentum space discrete approximation
\begin{eqnarray}
  \label{eq:continuous-position}
  f(\mathbf{x}) \propto \sum_{\lbrace s_i\rbrace} e^{-ip_{\mathbf{s}} \mathbf{x}} \langle{\mathbf{s}|\bm{\mathcal{F}} |f^{(n)}}\rangle.
\end{eqnarray}

We can also use the finite difference method to perform the interpolation. We can reconstruct the $(n+1)$-qubit function from the $n$-qubit one by dividing the spatial discretization step $\Delta x_i$ by two, and approximating the middle points as
\begin{align}
\begin{split}
  f(\mathbf{x} + \Delta \mathbf{x}/2) &\approx f(\mathbf{x}) + \frac{\Delta \mathbf{x}}{2} \nabla f(\mathbf{x} + \Delta  \mathbf{x}/2
  ) \\
                                      &\approx f(\mathbf{x}) + \frac{f(\mathbf{x} + \Delta\mathbf{x})-f(\mathbf{x})}{2},
\end{split}
\end{align}
where $\Delta \mathbf{x} = (\Delta x_1,\dots, \Delta x_d)$. Since we have focused on the finite differences approximation of the derivative, we will use this interpolation technique.

\section{Problem bond dimension}
\label{app:bond-dimension}

\begin{figure}[t]
  \centering
  \includegraphics[width=1\linewidth]{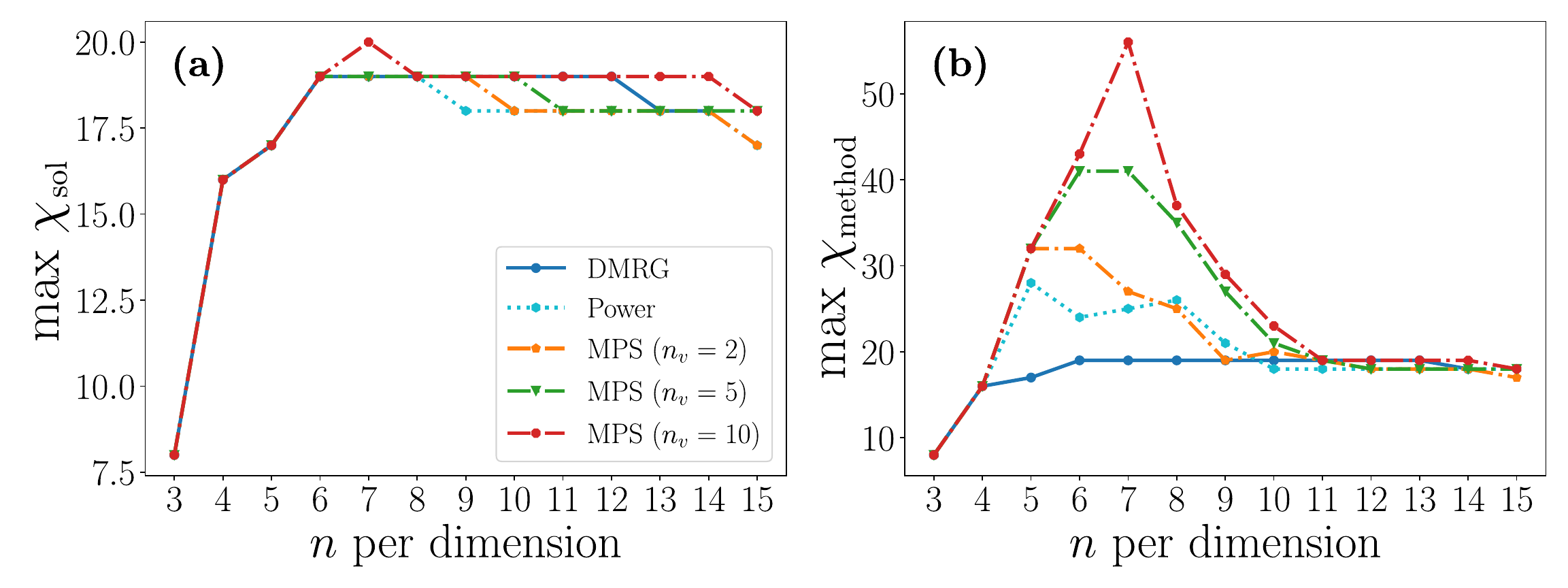}
  \caption{Scaling of the maximum bond dimension with the number of qubits $n$ per dimension for the 2D squeezed harmonic oscillator interpolation experiments. (a) Maximum bond dimension of the solution for each method $\chi_\text{sol}$, (b) maximum bond dimension reached for each method $\chi_\text{method}$.\label{fig:max_D}}
\end{figure}

In MPS methods, the bond dimension $\chi$ plays a key role in the trade-off between efficiency and accuracy. The bond dimension of the MPS is related to its memory needs, and hence it scales similarly with the number of qubits $n$ per dimension as the memory (Fig.\ \ref{fig:squeezed_ho_int}(b)). When using interpolation, the methods converge to the solution, leading to similar bond dimensions for all MPS methods for the final solution (Fig.\ \ref{fig:max_D}(a)). We observe that even if the maximum bond dimension fluctuates for the different algorithms (Fig.\ \ref{fig:max_D}(b)), all converge to a similar final bond dimension related to the entanglement structure of the function that represents the solution of the PDE.

\section{Extension of the global optimization to other PDEs} \label{app:source-PDEs}

The global optimization methods can solve another type of PDEs as long as we can rewrite them in the form of a cost function that will lead to the solution. A possible application is the resolution of PDEs with a source term $g(x)$,
\begin{eqnarray}\label{eq:source PDE}
  Df(x) = g(x), \quad f(x), g(x) \in \mathbb{C}^N.
\end{eqnarray}
We solve \eqref{eq:source PDE} by minimizing the cost functional $C[f]$,
\begin{eqnarray} \label{eq:cost source PDE}
  C[f] = || Df(x)-g(x)||^2.
\end{eqnarray}
The gradient descent algorithm\ \ref{sec:gradient-descent} then follows the optimization path
\begin{eqnarray}\label{eq:descent source eq}
  f_{n+1} = f_n + \Delta\beta D^\dagger(Df- g) = f_n + \Delta\beta D^\dagger w,
\end{eqnarray}
and using the optimum $\Delta\beta$ for each step,
\begin{eqnarray} \label{eq:step source eq}
  \Delta\beta_\mathrm{opt} = - \frac{\langle w | D D^\dagger | w\rangle}{\langle w | D D^\dagger D D^\dagger |w\rangle}.
\end{eqnarray}

The efficient resolution of PDEs with source terms is key, due to the multiple applications of such equations. Some important source PDEs are Poisson's equation and the heat equation, which have many applications beyond their original use, as many models can be reduced to them. An interesting example is the use of the heat equation in finance, as the Black-Scholes equation\ \cite{Black1973} can be expressed in terms of it.
\bibliography{my_bibliography}
\end{document}